\documentstyle[12pt,aaspp4]{article}

\def\linebreak{\hfil\break}

\def\etal{{\it et al}. }
\def\degree{\ifmmode {^\circ}\else {$^\circ$}\fi}
\def\mum{\ifmmode {\rm \mu {\rm m}}\else $\rm \mu {\rm m}$\fi}
\def\arcsec{\ifmmode ^{\prime \prime}\else $^{\prime \prime}$\fi}

\def\inch{\ifmmode ^{\prime \prime}\else $^{\prime \prime}$\fi}
\def\arcmin{\ifmmode ^{\prime}\else $^{\prime}$\fi}

\def\msun{\ifmmode {\rm M_{\odot}}\else $\rm M_{\odot}$\fi}
\newbox\grsign \setbox\grsign=\hbox{$>$} \newdimen\grdimen \grdimen=\ht\grsign
\newbox\simlessbox \newbox\simgreatbox
\setbox\simgreatbox=\hbox{\raise.5ex\hbox{$>$}\llap
     {\lower.5ex\hbox{$\sim$}}}\ht1=\grdimen\dp1=0pt
\setbox\simlessbox=\hbox{\raise.5ex\hbox{$<$}\llap
     {\lower.5ex\hbox{$\sim$}}}\ht2=\grdimen\dp2=0pt
\def\simgreat{\mathrel{\copy\simgreatbox}}
\def\simless{\mathrel{\copy\simlessbox}}

\begin{document}

\pagestyle{empty}
\singlespace

\centerline{\Large {\bf Accretion in the Early Kuiper Belt }}
\vskip 2ex
\centerline{\Large {\bf II. Fragmentation}}
\vskip 4ex
\centerline{Scott J. Kenyon}
\centerline{Harvard-Smithsonian Center for Astrophysics}
\centerline{60 Garden Street, Cambridge, MA 02138}
\centerline{e-mail: skenyon@cfa.harvard.edu}
\vskip 1ex
\centerline{and}
\vskip 1ex
\centerline{Jane X. Luu}
\centerline{Leiden Observatory}
\centerline{PO Box 9513, 2300 RA Leiden, The Netherlands}
\centerline{e-mail: luu@strw.leidenuniv.nl}
\vskip 3ex
\centerline{to appear in}
\centerline{{\it The Astronomical Journal}}
\centerline{July 1999}

%
%
\singlespace

\begin{abstract}

We describe new planetesimal accretion calculations in the Kuiper Belt
that include fragmentation and velocity evolution.  All models produce 
two power law cumulative size distributions, $N_C \propto r^{-2.5}$ 
for radii $\lesssim$ 0.3--3 km and $N_C \propto r^{-3}$ for radii 
$\gtrsim$ 1--3 km. The power law indices are nearly independent of 
the initial mass in the annulus, $M_0$; the initial eccentricity of 
the planetesimal swarm, $e_0$; and the initial size distribution 
of the planetesimal swarm.  The transition between the two power 
laws moves to larger radii as $e_0$ increases.  The maximum size 
of objects depends on their intrinsic tensile strength, $S_0$; Pluto 
formation requires $S_0 \gtrsim$ 300 erg g$^{-1}$.  The timescale 
to produce Pluto-sized objects, $\tau_P$, is roughly proportional 
to $M_0^{-1}$ and $e_0$, and is less sensitive to other input 
parameters.  Our models yield $\tau_P \approx$ 30--40 Myr for
planetesimals with $e_0 = 10^{-3}$ in a Minimum Mass Solar Nebula.  
The production of several `Plutos' and $\sim 10^5$ 50 km radius 
Kuiper Belt objects leaves most of the initial mass in 0.1--10 km 
radius objects that can be collisionally depleted over the age 
of the solar system.  These results resolve the puzzle of large 
Kuiper Belt objects in a small mass Kuiper Belt.

\end{abstract}

\section{INTRODUCTION}
\pagestyle{plain}

Recent discoveries of slow-moving objects beyond the orbit of Neptune
have radically changed our understanding of the outer solar system.   
These observations have revealed a large population of Kuiper Belt 
objects (KBOs) in orbits with semi-major axes of 39--45 AU (\cite{jew93}, 
1995; \cite{ir95}; \cite{wil95}; \cite{jew96}; \cite{luu97};
\cite{gla97}; \cite{gla98}).  KBOs with reliable orbits have a 
cumulative size distribution that follows $N_C \propto r^{-q}$, 
with $q = 3 \pm 0.5$ (\cite{jew98}; \cite{luu98}).  The estimated 
population of $\sim 10^5$ KBOs with radii $r \gtrsim$ 50 km 
indicates a total mass, $M_{KBO} \approx 0.1 M_E$\footnote{1 $M_E 
= 6 \times 10^{27}$ g}, for reasonable assumptions about the albedo 
and distance distribution.  

The small mass in KBOs is a problem for current planet formation 
theories.  In most theories, small planetesimals in the solar nebula 
grow by collisional accumulation (e.g., \cite{saf69}; 
\cite{gre78}, 1984; \cite{wet89}, 1993; \cite{spa91}).  This growth
eventually produces one or more `cores' that accumulate most, if not 
all, of the solid mass in an annular `feeding zone' defined by balancing 
the gravity of the growing planetesimal with the gravity of the Sun and 
the rest of the disk.  Large cores with masses of 1--10 $M_E$ can 
also accrete gas from the feeding zone (\cite{pol84}).  Applied to 
the inner solar system, this model generally accounts for the masses
of the terrestrial and gas giant planets (e.g., \cite{pol96}, \cite{we97b};
but see \cite{bos97}).  At 30--50 AU, however, the timescale to produce 
planet-sized, $\sim$ 1000~km, objects exceeds the disk lifetime, 
$\simless$ 100 Myr, unless the mass of the outer disk is 
$\sim 10^2$--$10^3~M_{KBO}$ (e.g., \cite{fer81}, 1984; \cite{ip89}; 
\cite{ste95}, 1996; \cite{st97a}).  The presence of large KBOs in a 
small mass Kuiper Belt is thus a mystery.

In Kenyon \& Luu (1998, hereafter KL98), we began to address this issue
by considering planetesimal growth in a single annulus at 35 AU.  We 
showed that calculations including accretion and velocity evolution
naturally produce several ``Plutos'' with radii exceeding 1000 km and
numerous 50 km radius KBOs on timescales of $\tau \sim$ 20--200 Myr for 
a wide range of initial conditions in plausible solar nebulae.  These 
timescales indicate that Pluto can form in the outer solar system 
in parallel with the condensation of the outermost large planets.

In this paper, we extend KL98 by adding fragmentation to our
planetesimal evolution code.  The code generally matches published
calculations at 1 AU (Wetherill \& Stewart 1993, hereafter WS93)
and at 40 AU (Davis \& Farinella 1997).  Our numerical results 
demonstrate that fragmentation and velocity evolution damp 
runaway growth to provide a self-limiting mechanism for the 
formation of KBOs and Pluto in the Kuiper Belt.  
These calculations produce several Plutos and at least $10^5$ 
50 km radius KBOs on timescales of 2--100 Myr in annuli with modest 
surface densities, 2.0--0.14 g cm$^{-2}$, of solid material at 35 AU.  
Our analysis sets a lower limit on the intrinsic strength of KBOs, 
$\sim$ 300 erg g$^{-1}$, and indicates that the initial size distribution, 
the initial eccentricity of the planetesimal swarm, and the details of 
the fragmentation algorithm have little impact on the resulting size 
distribution of KBOs.  

Our results appear to resolve the mystery of large KBOs in a small mass
Kuiper Belt.  Planetesimal evolution at 35--50 AU in a Minimum Mass
Solar Nebula (\cite{hay81}) naturally produces large KBOs in numbers close 
to those currently observed.  Most of the disk mass ends up in smaller 
objects, 0.1--10 km, that can be collisionally depleted over the age of 
the solar system.  This depletion rate depends on the intrinsic strength 
and eccentricity distribution of KBOs (\cite{ste96}; \cite{dav97}; 
\cite{st97a}, 1997b).  Future observations can place better constraints 
on these physical parameters and provide additional tests of our 
interpretation of KBO formation.

We outline the fragmentation model and tests in Sec. 2,
describe calculations for the Kuiper Belt in Sec. 3, and
conclude with a discussion and summary in Sec. 4.
The Appendix contains a complete description of the
fragmentation algorithms and updates of the coagulation
code from KL98.

\section{The Model}

As in KL98, we adopt Safronov's (1969) particle-in-a-box method, where 
planetesimals are treated as a statistical ensemble of masses with a 
distribution of horizontal and vertical velocities about a Keplerian orbit.  
Our calculations begin with a differential mass distribution $n(m_i$) 
in a single accumulation zone centered at a heliocentric distance
$a$ with an inner radius at $a - \Delta a/2$ and an outer radius 
at $a + \Delta a/2$.  The horizontal and vertical velocity dispersions 
of these bodies are $h_i(t)$ and $v_i(t)$.  We approximate the continuous 
distribution of particle masses with discrete batches having particle 
populations $n_i(t)$ and total masses $M_i(t)$ (WS93).  The average mass 
of a batch, $m_i(t)$ = $M_i(t) / n_i(t)$, changes with time as collisions 
add and remove bodies from the batch.  This procedure naturally 
conserves mass and allows a coarser grid than calculations with 
fixed mass bins (\cite{wet90}; \cite{ws93}).

To evolve the mass and velocity distributions in time, we solve the
coagulation and energy conservation equations for an ensemble of
objects with masses ranging from $\sim 10^{7}$~g to $\sim 10^{26}$ g.  
The Appendix of KL98 describes our treatment of accretion and 
velocity evolution, and compares numerical results with analytic 
solutions for standard test cases.  In this paper, we add fragmentation 
to the problem. We adopt a simple treatment of collision outcomes,
following Greenberg \etal (1978; see also \cite{dav85}, 1994; WS93):

\vskip 4ex
\noindent
1. Mergers -- two bodies collide and merge into a single object
with no debris;

\noindent
2. Cratering -- two bodies merge into a single object but produce
debris with a mass much smaller than the mass of the merged object;

\noindent
3. Rebounds -- two bodies collide and produce some debris but do 
not merge into a single object; and

\noindent
4. Disruption -- two bodies collide and produce debris with a mass
comparable to the mass of the two initial bodies.
\vskip 4ex

We consider two algorithms for treating the cratering and disruption 
of planetesimals.  Both the WS93 and the Davis \etal (1984) algorithms 
estimate the amount of debris $m_{f,ij}$ produced from impacts with 
velocities exceeding a threshold velocity $V_f$. In general, $m_{f,ij}$ 
scales with the impact energy.  WS93 assume that the debris has the same 
relative velocity $V_{ij}$ of the colliding bodies;
Davis \etal (1984) assume that the debris receives a fixed fraction
$f_{KE}$ of the impact kinetic energy. In both cases, we adopt a 
coefficient of restitution $c_R$ to allow for rebound collisions that 
produce debris but no single merged body. We follow Greenberg \etal 
(1978) and adopt separate values of $c_R$ for collisions with ($c_2$)
and without ($c_1$) cratering. The Appendix describes these algorithms 
in more detail.

To test our fragmentation procedures, we attempt to duplicate WS93's 
calculations of planetary embryo formation at 1 AU.  Their model
begins with 8.33 $\times~10^8$ planetesimals having radii of 8 km and 
a velocity dispersion of 4.7 m s$^{-1}$ relative to a Keplerian orbit 
(Table 1; see also Table 1 of \cite{ws93}).  Table 2 summarizes our 
results using the WS93 initial conditions with mass spacing factors 
of $\delta \equiv m_{i+1}/m_i$ = 1.12, 1.25, 1.4, and 2.0 between 
successive mass batches.  We adopt the analytic cross-sections of 
Spaute \etal (1991), which yield results identical to the numerical 
cross-sections of WS93 (\cite{kl98}).  Figure 1 shows our reproduction 
of WS93 for $\delta = 1.25$.  This model produces twelve 3--7 
$\times~10^{25}$ g objects with velocity dispersions of 60--100 m s$^{-1}$ 
in 1.15 $\times~10^5$ yr.  In contrast, WS93 produce seven 1--3 
$\times~10^{26}$ g objects (see Figures 1--3 in WS93).  Despite this 
factor of two difference in the total mass contained in the most 
massive bodies, our calculation has the same broad ``plateau'' in 
the cumulative number distribution $N_C$ at log $m_i$ = 24--26 and 
a similar power law dependence, $N_C \propto m_i^{-1}$, at log $m_i$ 
= 21--23.  Our ``fragmentation tail'' at log $m_i$ = 7--18 has the 
standard power law dependence, $N_C \propto m_i^{-0.8}$ (\cite{doh69}),
at large masses but flattens out more than the WS93 result at small masses.

The evolution of particle velocities in our calculations generally 
agrees with WS93 (Figure 2; see Figures 2--3 of WS93).  
All velocities increase monotonically with time due to viscous stirring.  
The velocities of large bodies grow very slowly, because 
dynamical friction transfers their kinetic energy to small bodies.
The model maintains a nearly constant ratio of vertical to 
horizontal velocity, $v_i/h_i \approx$ 0.53, for all but the most 
massive bodies, which have $v_i/h_i < 0.5$ (Figure 2, right panel).  
This result agrees with previous calculations (Barge \& Pellat 
1990, 1991; \cite{hor85}).  At late times, our velocities for 
small bodies, $h_i \approx$ 200 m s$^{-1}$ at $m_i \sim 10^{7}$ g, 
are roughly a factor of two larger than those of WS93.  
Velocities for large bodies, $h_i \approx$ 100 m s$^{-1}$ 
at $m_i \sim 10^{25}$ g, roughly equal those of WS93.

The higher velocities of our calculation lead to additional mass
loss from gas drag and fragmentation.  Gas drag removes material 
from the annulus; fragmentation produces bodies with masses less 
than the minimum mass, $\sim 10^7$ g, of our numerical grid. We
typically lose $\sim$ 25\% of the initial mass to cratering and 
catastrophic fragmentation and another 20\%--25\% to gas drag.  
WS93 lost a comparable amount of mass to fragmentation but only 
$\sim$ 5\% to gas drag.  

The agreement between our results and those of WS93 depends on 
the mass spacing factor $\delta$.  For $\tau \lesssim$ 30,000 yr, 
large $\delta$ models take longer to produce 1000--2000 km 
objects than small $\delta$ models.  The `lag' relative to our
$\delta$ = 1.12 model increases from 1\%--2\% for $\delta$ = 1.25 
up to $\sim$ 10\% for $\delta$ = 2 (see also \cite{oht88}; 
\cite{oht90}; \cite{wet90}; \cite{kol92}; KL98). 
The poor resolution of $\delta \ge 1.4$ models delays the production of 
massive bodies that `runaway' from the rest of the mass distribution. 
In WS93, most runaway bodies are also `isolated' bodies that do not 
interact with one another but do interact with the rest of the mass 
distribution (see WS93 and the Appendix).  Delays in the production 
of isolated, runaway bodies lead to 10\%--20\% increases in the 
velocity dispersion of all bodies.  Larger velocities reduce 
gravitational focusing factors and slow the growth of the largest 
bodies.  The mass of the largest object thus increases from 
7--8 $ \times~10^{25}$ g for $\delta$ = 1.25--1.4 to 
$9.5 \times 10^{25}$ g for $\delta$ = 1.12.  This trend suggests 
that calculations with $\delta$ = 1.01--1.10, as in WS93, would 
improve the agreement between our results and those of WS93.

The coagulation results also depend on the treatment of low velocity 
collisions.  Our standard expressions for long-range gravitational 
interactions fail when the velocity dispersion is close to the 
Hill velocity, $v_H$ (\cite{ida90}; \cite{bar91}).  We use Ida's 
(1990) $n$-body calculations and adopt his simple expressions to 
derive the velocity evolution for collisions in the low velocity 
limit, $V_{ij} < V_{lv} v_H$, where $V_{lv}$ = 2--5 is a constant 
(see the Appendix).
This change slows down the velocity evolution, because it introduces
a lower limit to the timescales for dynamical friction and the
inclination component of viscous stirring (\cite{ida90}). We adopt
$V_{lv}$ = 2, following WS93, for 1 AU calculations.  Unlike WS93,
however, the small bodies in our calculation do not quite reach the 
low velocity limit and increase in velocity throughout the evolution.
A modest increase in our low velocity limit, $V_{lv}$ = 3.5, leads 
to better agreement between the velocity behavior of our models and 
that of WS93.

To test our procedures further, we repeat the 1 AU calculations using
the Davis \etal (1984) fragmentation algorithm.  Table 3 lists our
results for $f_{KE}$ = 0.1 and $\delta$ = 1.25, 1.4, and 2.0.  These
calculations yield solutions similar to our WS93 models.  Calculations 
with smaller $\delta$ produce larger bodies at earlier times than 
models with large $\delta$ (Table 3).   The mass of the largest 
object increases from $m_i \approx 7 \times 10^{25}$ g for $\delta$ = 
2.0 to $m_i \approx 1.2 \times 10^{26}$ g for $\delta$ = 1.4 to
$m_i \approx 1.8 \times 10^{26}$ g for $\delta$ = 1.25.  The velocity 
dispersions of the small bodies are a factor of 2--3 larger at $\tau 
\lesssim$ 30,000 yr because they receive a larger fraction of the 
fragmentation energy than in the WS93 algorithm.  This trend reverses 
for $\tau \gtrsim$ 50,000 yr, because the Davis \etal algorithm produces 
more debris for collisions between large objects than the WS93 algorithm.
These large objects have small velocity dispersions, so the velocity 
dispersions of the small bodies increase more slowly than in the WS93 
model.  The mass lost from fragmentation, $\sim$ 5\% of the initial mass,
and gas drag, $\sim$ 20\% of the initial mass, is also smaller.

We conclude that our coagulation code with fragmentation and velocity
evolution reproduces `standard' calculations.  The results in Tables
2--3 generally agree with published results, despite some differences 
in the evolution of the velocity dispersion and the radius of the 
largest object.  These discrepancies probably result from subtle 
differences in the implementation of the algorithms.  None seem 
significant, because the major results do not depend on the input 
parameters (see also WS93): all calculations produce several objects 
with $m_i \sim 10^{26}$ g that contain most of the remaining mass 
in the annulus.

\vfill
\eject

\section{Kuiper Belt Calculations}

\subsection{Starting Conditions}

As in KL98, we rely on observations of other stellar systems and 
solar nebula models to choose appropriate initial conditions for our 
Kuiper Belt  calculations.  Recent observations indicate lifetimes of $\sim$ 
5--10 Myr for typical gaseous disks surrounding nearby 
pre--main-sequence stars and for the solar nebula (\cite{sar93}; 
\cite{rus96}; \cite{har98}).  In our solar system, Neptune formation 
places another constraint on the KBO growth time, because Neptune excites 
KBOs through gravitational perturbations.  Recent calculations suggest
Neptune can form in 5--100 Myr (\cite{ip89}; \cite{lis96}; \cite{pol96}).
Once formed, Neptune inhibits KBO formation at 30--40 AU by increasing 
particle random velocities on timescales of 20--100 Myr (\cite{hol93}; 
\cite{dun95}; \cite{mor97}).  We thus adopt 100 Myr as an upper limit 
to the KBO formation timescale at 30--40 AU.

Our starting conditions for KBO calculations are similar to KL98.
We consider an annulus centered at 35 AU with a width of 6 AU.  
This annulus can accommodate at least 10--100 isolated bodies with 
$m_i \simgreat 10^{24}$ g for $e \le 0.01$.  Instead of the single 
starting radius used in KL98, the present calculations begin with $N_0$ 
bodies in a size distribution with radii, $r_i$ = 1--80 m.  This 
initial population has a power law cumulative size distribution,
$N_C \propto r_i^{-q_0}$.  We usually adopt $q_0 = 3$; the final size 
distribution appears to depend very little on $q_0$, as discussed below. 
The planetesimals have a small initial eccentricity (\cite{mal95})
that is independent of size and an equilibrium ratio of inclination 
to eccentricity, $\beta_0 = \langle i_0 \rangle/\langle e_0 \rangle$ = 0.6 
(\cite{bar90}).  The mass density of each body is 1.5 g cm$^{-3}$.
To set the initial mass of the annulus, $M_0$, we extend the Minimum 
Mass Solar Nebula to the Kuiper Belt and integrate the surface density 
distribution for solid particles, $\Sigma = \Sigma_0 (a/a_0)^{-3/2}$, 
across the 6 AU annulus.  The dust mass is then $M_{min} \approx 
0.25~\Sigma_0~M_E$ at 32--38 AU for $a_0$ = 1~AU.  Most Minimum Mass 
Solar Nebula models have $\Sigma_0$ = 30--60 g cm$^{-2}$ (\cite{wei77}; 
\cite{hay81}; \cite{bai94}), which sets $M_{min} \approx$ 7--15 $M_E$.
We consider models with $M_0$ = 1--100 $M_E$ to allow for 
additional uncertainty in $\Sigma_0$.  Table 1 compares input 
parameters for Kuiper Belt models with initial conditions 
at 1 AU (see also \cite{ws93}).  

Our success criteria are based on available observations of KBOs.
The present day Kuiper Belt contains 
(a) at least one object (Pluto) with radius $\gtrsim$ 1000 km and
(b) $\sim$ 70,000 objects with radii exceeding 50 km 
(\cite{jew95}; \cite{jew96}, 1998).  A successful KBO calculation
must satisfy both observed properties in $\simless$ 100 Myr.  
We quantify these criteria by defining $r_{max}$ as the radius of
the largest object and $r_5$ as the radius where the cumulative 
number of objects is $N_C \ge 10^5$: a successful simulation has
$r_{max} \gtrsim$ 1000 km and $r_5 \simgreat$ 50 km at 
$\tau \lesssim$ 100 Myr.  To provide another characteristic 
of these models, we define $r_{95\%}$ such that 95\% of the 
mass is contained in objects with $r_i < r_{95\%}$.
In models with a long runaway growth phase, the largest objects contain
most of the mass and $r_{95\%} \approx r_5$.  Models with a limited 
runaway growth phase leave most of the mass in small objects, so 
$r_{95\%} \ll r_5$.  We end calculations with velocity evolution
when $r_{max}$ exceeds $\sim$ 1000 km.  To evaluate the dependence 
of runaway growth on fragmentation, we extend calculations without 
velocity evolution to 5000 Myr or to when $r_{max}$ exceeds $\sim$
2000--3000 km.

\subsection{Models Without Velocity Evolution}

To isolate how fragmentation changes the growth of KBOs,
we begin with constant velocity calculations of accretion.
The initial size distribution has $q_0 = 3$, $\delta$ = 1.4,
and a maximum initial radius of $r_0$ = 80 m. 
The initial velocities of $h_i$ = 4 m s$^{-1}$ and $v_i$ = 
2.1 m s$^{-1}$ correspond to an equilibrium model with $e_0 
= 10^{-3}$ (\cite{hor85}).  We use the Davis \etal (1985) 
fragmentation algorithm and assume that the collision fragments 
receive a kinetic energy per unit mass equal to one-half of 
the square of the relative velocity of the colliding bodies, 
$V_{ij}$.  The bodies are strong, with a tensile strength 
$S_0 = 2 \times 10^6$ erg g$^{-1}$. 
Tables 1 and 4 summarize the initial conditions and results for models 
with $M_0$ = 1--100 $M_E$, $e_0$ = $10^{-3}$, and the coefficients 
of restitution, ($c_1$,$c_2$) = ($10^{-5}$,$10^{-5}$) and 
($10^{-2}$,$10^{-3}$). 

Figure 3 shows how $N_C$ evolves with time for $M_0$ = 10 $M_E$, 
$e_0 = 10^{-3}$, and ($c_1$,$c_2$) = ($10^{-5}$,$10^{-5}$).  
The low coefficients of restitution eliminate rebound collisions. 
Roughly half of the most massive objects experiences at least one 
collision by $\tau \approx$ 16~Myr, when the 18 largest bodies have 
$r_i \approx $ 1~km.  These bodies reach $r_i \sim$ 10 km at 
$\tau \approx$ 135 Myr and grow to 100~km sizes at 255 Myr.  
The growth rate then increases considerably due to gravitational 
focusing.  Runaway growth ensues.  The largest planetesimals 
reach $r_{max} \approx$ 200~km at $\tau \approx$ 265~Myr; 
$r_{max}$ exceeds 1000~km only 11~Myr later.  At $\tau \approx$ 280~Myr,
a single runaway body with $r_{max} \approx$ 2000~km begins to sweep up 
lower mass planetesimals and contains nearly all of the mass in 
the annulus a few Myr later.

The cumulative size distribution of small objects with $r_i \lesssim$ 
100 m slowly approaches a power law with $q = 2.25$ throughout the 
evolution.  The distribution is shallow, $q < 2$, for $\tau \lesssim$
20 Myr as growth produces many 50--500 m objects.  The distribution 
steepens as the largest bodies grow past 1 km and settles at 
$q = 2.25$ for $\tau \gtrsim$ 150 Myr.   This distribution
is steeper than the theoretical limit of $q = 2.5$ (\cite{doh69}; 
\cite{wil94}), because growth and fragmentation never reach equilibrium. 

Calculations with rebound collisions yield similar results. With larger
coefficients of restitution, ($c_1$,$c_2$) = ($10^{-2}$,$10^{-3}$), 
rebound collisions occur for $r_i \lesssim$ 5 m.  These collisions 
produce fragments but no mergers into larger bodies.  As a result, the size distribution is initially very steep for 
$r_i \lesssim$ 5 m and very shallow for $r_i \gtrsim$ 5 m (Figure 4).  
As the largest objects grow from 1 km to 10 km at $\tau$ = 16--135 Myr,
bodies with $r_i \lesssim$ 5 m are swept up by the large bodies 
and replaced by collision fragments.  This process smooths out the 
size distribution at $r_i \sim$ 1--20 m, although the slope still 
changes at $r_i \sim$ 5 m.  

The evolution of the largest bodies is unaffected by rebounds.  
These objects reach sizes of 100 km only slightly later than models 
without rebounds, 258 Myr {\it vs} 255 Myr, and then begin to runaway from 
lower mass planetesimals.  Several objects reach radii of 1000 km in 
another $\sim$ 20 Myr and then begin to consume all of the material 
left in the annulus.

Fragmentation is not important in any of these low eccentricity 
calculations.  Catastrophic fragmentation produces no mass loss.
because the velocities remain artificially low. These models lose 
only 1\%--3\% of their initial mass due to cratering.  The 
timescale to grow into a 1000 km object is nearly identical to 
models without fragmentation (see KL98): Pluto forms in 
$\tau_P \approx$ 276--280 Myr ($M_0/\rm 10 ~ M_E$)$^{-1}$ for
$r_0$ = 80 m.  Calculations with larger $r_0$ should have smaller 
growth times (see KL98).

\subsection{Models With Limited Velocity Evolution}

To understand how fragmentation changes the velocity dispersions
of colliding planetesimals, we consider calculations with `limited'
velocity evolution.  As in Davis \etal (1985, 1994), we assume 
that collision fragments receive a fixed fraction $f_{KE}$ 
of the center-of-mass impact energy, but consider no other 
changes to the kinetic energy of the bodies.  For most collisions, 
this assumption produces fragments with velocities larger than the 
relative collision velocity $V_{ij}$.  To conserve kinetic energy, 
the merged bodies have a lower velocity dispersion after the collision.
This redistribution of velocity mimics dynamical friction and 
should lead to more rapid growth than models with no velocity 
evolution.  Table 5 summarizes results for $e_0$ = $10^{-3}$, 
$S_0 = 2 \times 10^6$ erg g$^{-1}$, and coefficients of restitution, 
($c_1,c_2$) = ($10^{-2},10^{-3}$), for our standard mass distribution 
with $q_0 = 3$, $\delta$ = 1.4, and $r_0$ = 80 m.

Figure 5 shows how $N_C$ evolves with time for $M_0$ = 10 $M_E$.
During the first 10--20 Myr, the size distribution has two power 
laws, $q \approx 4$ for $r_i \lesssim$ 5 m and $q \approx 1.25$ 
for $r_i \gtrsim$ 5 m.  This break in the slope is due 
to rebounds, which prevent growth of the smallest objects.  As the
largest objects grow to sizes of 1--10 km, collision debris adds 
to the population of 5~m objects.  The slope of the size 
distribution is then $q \approx 2.5$ for $r_i \lesssim$ 100 m.
For large objects, $q$ approaches 2.5 once $r_{max}$ exceeds 100 km.  
Once $r_{max} \gtrsim$ 1000 km, the complete number distribution 
follows a power law with $q = 2.5$, except for a small kink at 
$r_i \approx$ 1 km whose amplitude decreases with time.

Limited velocity evolution eventually produces larger objects
on shorter timescales than models without velocity evolution.
In the first 20--40 Myr, collisions between the largest bodies,
$r_i \approx$ 100--500 m, add material with high velocity dispersion
to batches with $r_i \sim$ 1--10 m.  The horizontal velocity dispersions 
of these low mass objects increase from 4 m s$^{-1}$ to 10--30 m s$^{-1}$
in 17 Myr and reach a roughly constant value of $\sim$ 20 m s$^{-1}$ 
after 200 Myr (Figure 6).  This large velocity dispersion initially 
slows down the growth of the largest objects relative to models with 
no velocity evolution. It takes this model $\sim$ 1 Myr longer to
produce objects with $r_i \sim$ 1 km, but these objects then have
velocities $\sim$ 1 m s$^{-1}$ smaller than the initial velocity 
of the planetesimal swarm (see Figure 6).  This smaller velocity
dispersion enhances the growth of the largest objects once gravitational
focusing becomes important. The model produces 10 km objects in 
120 Myr and 100 km objects in 202 Myr.  This rapid evolution leads 
to runaway growth and the production of a few 1000+ km objects after
another 15 Myr. 

Figure 7 compares $r_{max}$ in models with no velocity evolution 
(thin solid lines) and models with limited velocity evolution 
(thick solid lines).  The annuli have initial masses $M_0$ = 1, 
10, and 100 $M_E$.  These models are nearly indistinguishable for 
$r_{max} \lesssim$ 10 km. The limited velocity evolution models 
then produce larger objects on shorter timescales. These models
do not have a mass-dependent velocity evolution such as viscous 
stirring or dynamical friction, so the timescale to produce 1000 km 
objects is still inversely proportional to the mass in the annulus, 
$\tau_P$ = 216 Myr $(M_0/10 ~ M_E)^{-1}$ for $e_0$ = $10^{-3}$.

Other than the accelerated growth rate at late times, there is little 
difference between models with limited velocity evolution and models
with no velocity evolution (see Tables 4--5).  Both sets produce 
comparable numbers of large KBOs, $r_i \gtrsim$ 100 km, that 
contain most of the mass in the annulus when the first Pluto 
forms.  These models also lose $\sim$ 1\%--3\% of their initial mass 
due to cratering. The mass loss depends solely on the total kinetic
energy of all the bodies, which is a constant in these models.
Catastrophic fragmentation still produces no mass loss, because
we assume relatively strong objects with $S_0 = 2 \times 10^6$ 
erg g$^{-1}$. We describe models with weaker bodies in the next section.  

\subsection{Models with Velocity Evolution}

We now consider a complete coagulation calculation with fragmentation
and velocity evolution.  In addition to velocity redistribution from
fragmentation, these models include (i) gas drag, 
(ii) dynamical friction and viscous stirring from 
long-range (elastic) collisions, and
(iii) dynamical friction and viscous stirring from 
short-range (inelastic) collisions (see the Appendix of KL98).
To approximate the disappearance of the solar nebula on reasonable
timescales, we assume that the gas density decays exponentially 
with time, $\rho_g(t) = \rho_{g,0}~e^{-t/\tau_g}$ where 
$\rho_{g,0} = 1.18 \times 10^{-9}~(a/1 AU)^{-21/8}~(M_0/4 M_E)$
g cm$^{-3}$ (WS93 and references therein).  The influence of 
gas drag on the planetesimals thus decreases with an $e$-folding 
time of $\tau_g$, which we set at $\tau_g$ = 10 Myr (\cite{har98}).
We adopt $\delta$ = 1.4, $q_0 = 3$, and $r_0$ = 80 m. The initial 
velocities are $h_i$ = 4.0 ($e_0/10^{-3}$) m~s$^{-1}$ and 
$v_i$ = 2.1 ($e_0/10^{-3}$) m~s$^{-1}$. Tables 6--7 summarize 
the initial conditions and results for models with 
$M_0$ = 1--100 $M_E$ and $e_0$ = $10^{-4}$ to $10^{-2}$.

\subsubsection{A Standard Model with $e_0 = 10^{-3}$
and $S_0 = 2 \times 10^6$ erg g$^{-1}$}

Figure 8 shows how $N_C$ and $h_i$ evolve for $M_0$ = 10 $M_E$,
$e_0 = 10^{-3}$, and $S_0 = 2 \times 10^6$ erg g$^{-1}$. 
Inelastic collisions act rapidly to circularize the orbits of 
objects with $r_i \gtrsim$ 5 m, which decrease in velocity from 
$h_i$ = 4 m s$^{-1}$ to $h_i$ = 1--2 m s$^{-1}$ in 5 Myr. Larger 
bodies have less frequent collisions; their orbits circularize on longer
timescales.  Small bodies with $r_i \lesssim$ 5~m collide frequently,
but rebounds prevent these bodies from circularizing their orbits.
Instead, collision fragments increase the velocity dispersion of the 
lowest mass bodies.  These processes produce an inverted velocity 
distribution, where bodies with $r_i \approx$ 5 m have smaller
velocities than objects with $r_i \lesssim$ 5 m and $r_i \gtrsim$
10 m. 

The steady damping of particle velocities enhances growth of the
larger bodies relative to calculations with limited velocity 
evolution.  The largest objects grow slowly for 8.5 Myr, when there 
are $\sim$ 500 objects with $r_i$ = 1 km. The size distribution then
has three main features: 
(i) a pronounced fragmentation tail with a modest velocity dispersion, 
$h_i \approx$ 1--3 m s$^{-1}$;
(ii) a transition region with pronounced kink in the size distribution; 
and (iii) a group of rapidly growing bodies with low velocities,
$h_i$ = 0.02 m s$^{-1}$ (Figure 8; left panel).  These features remain 
prominent as $r_{max}$ increases from 1 km to 100 km in only 3.5 Myr.  

During runaway growth, dynamical friction and viscous stirring begin 
to increase particle velocities (Figure 8; right panel).  This phase
ends when $r_{max}$ reaches $\sim$ 300 km at 15 Myr.  The velocity 
dispersions of the small to intermediate mass bodies are then 
large enough, $\sim$ 10~m s$^{-1}$, to reduce gravitational 
focusing factors considerably.  The largest objects enter a 
steady growth phase, where their radii grow slowly with time. 
There are 2 Charon-sized objects with 
$r_i \gtrsim$ 500 km at 19 Myr; 50 ``Charons'' at 25 Myr, and
$\sim$ 150 Charons at 36.5 Myr when the first ``Pluto'' appears. 
The size distribution then closely follows a power law with 
$q = 2.5$ for $r_i \lesssim$ 30~m and a steeper power law with 
$q = 3$ for $r_i$ = 1--1000 km.  The modest velocity dispersion 
of the largest objects, $h_i \approx$ 0.5 m s$^{-1}$, maintains 
steady growth for another 60 Myr: 
4 objects have $r_i$ = 1450 km at 50 Myr;
8 objects have $r_i$ = 2000 km at 100 Myr.

Figure 9 illustrates how the largest objects grow as a function of $M_0$.
All models begin with slow growth, where inelastic collisions reduce 
the velocity dispersions of the intermediate mass objects. Slow growth 
lasts longer as $M_0$ decreases, because the growth rate depends on 
the collision rate.  Viscous stirring prolongs the slow growth phase 
of small $M_0$ models by counteracting collisional damping earlier 
in the evolution.  Slow growth thus produces more 1--2 km objects 
in models with smaller $M_0$.  Runaway growth eventually turns 1 km 
bodies into 100 km objects.  This phase ends when bodies with 
$r_i \lesssim$ 1 km have velocity dispersions exceeding 10--20 m s$^{-1}$.  
The large gravitational focusing factors that began runaway growth are
then smaller by several orders of magnitude.  The growth then returns 
to a steady phase where the largest bodies gradually approach radii 
of 1000 km.

In these models, cratering also acts to damp runaway growth.  As noted
in Sect. 3.3, collision debris increases the velocities of the
smallest objects.  Viscous stirring and dynamical friction enhance
this evolution as the largest objects grow beyond 1 km. The velocity 
dispersions of small objects increase from $\sim$ 20 m s$^{-1}$ at 
the end of runaway growth to $\sim$ 100 m s$^{-1}$ at 100 Myr.  
The largest objects then have $r_i \approx$ 2000 km.  
Collision debris produces $\sim$ 30\% of this increase; dynamical 
friction and viscous stirring are responsible for the rest.

Although all models with $e_0$ = $10^{-3}$ make at least one Pluto, 
only annuli with $M_0 \sim$ 10 $M_E$ meet both success criteria.  
In each calculation, $r_5$ steadily increases with time 
during the slow growth phase (Figure 10).  The number of 50 km radius 
KBOs increases dramatically during runaway growth, when gravitational 
focusing factors are large.  At the end of runaway growth, $r_5$ begins 
to decrease with time as the largest bodies try to separate themselves 
from the rest of the mass distribution.  This evolution is modest and 
short-lived, because the runaway growth phase ends quickly.  The number 
of 50 km radius KBOs then approaches a roughly constant value which
grows with increasing $M_0$.  Based on Figure 10, annuli with 
$M_0 \lesssim$ 3 $M_E$ produce too few 50 km radius KBOs compared 
with our success criterion; annuli with $M_0 \gtrsim$ 30 $M_E$ 
probably produce too many.  Annuli with 3 $M_E$ $\lesssim M_0 \lesssim$ 
30 $M_E$ can produce 1 or more Plutos and roughly $10^5$ KBOs in
$\lesssim$ 100 Myr.

This `standard model' has several important input parameters, 
including the initial eccentricity, the initial size distribution, 
and the intrinsic strength of the bodies.  To understand how 
the evolution depends on these parameters, we now consider 
variations on the standard model.  We begin with a discussion 
of $e_0$ and then describe models with different input 
strengths and size distributions.

\subsubsection{Models with $e_0 = 10^{-4}$ and $10^{-2}$ for
$S_0 = 2 \times 10^6$ erg g$^{-1}$}

The velocity dispersion of the planetesimal swarm affects growth
primarily through gravitational focusing.  Large bodies grow 
faster when velocity dispersions are small and gravitational 
focusing factors are large.  Despite small $e_0$, all of our
calculations begin in the ``high velocity regime,'' where 
gravitational focusing factors are near unity.  Collisional
damping and dynamical friction reduce velocity dispersions
to the low velocity regime where growth is rapid.  Planetesimals
with small $e_0$ can reach this regime first, so we expect that
smaller $e_0$ leads to more rapid growth.

Calculations with $e_0 = 10^{-4}$ and $10^{-2}$ confirm these 
general considerations.  In models with $e_0 = 10^{-2}$, the larger 
initial velocities reduce collisional damping compared to models
with $e_0 = 10^{-3}$.  Gravitational focusing factors thus
increase slowly.  Additional cratering debris also counteracts 
collisional damping and combines with dynamical friction and
viscous stirring to keep velocity dispersions large. As a result,
models with $e_0 = 10^{-2}$ experience a prolonged linear growth phase.
This phase is $\sim$ 10 times longer than for $e_0 = 10^{-3}$ models.
Slow growth ends when the largest objects reach the sizes needed
to produce modest gravitational focusing factors.  A short 
runaway growth phase eventually produces several Plutos.
The final size distribution has a larger fraction of mass in 
more massive objects, as indicated by a large value for $r_{95\%}$,
but has roughly comparable numbers of 50 km radius KBOs (Table 6).
These models also lose more mass to dust, $\sim$ 50\%, compared to 
the 1\%--3\% lost in models with $e_0 = 10^{-3}$.

Growth is rapid in models with $e_0 = 10^{-4}$, because 
collisional damping quickly reduces the velocity dispersion.  
This change does not have a dramatic effect on the growth of 
KBOs, because damping quickly reduces particle velocities 
to the low velocity limit.  The damping time is then 
independent of velocity.  Viscous stirring is also less 
effective. In our calculations, the growth times for 
$e_0 = 10^{-4}$ models are a factor of $\sim$ 2--10 smaller
than $e = 10^{-3}$ models (Table 6).  Models with 
$e_0 \lesssim$ a few $\times~10^{-5}$ probably have similar 
growth times, but we have not investigated this possibility 
in detail.

Figure 11 compares the size distribution near the end of our
calculations for three different values of $e_0$.  All 
models produce two power law size distributions, with
$q = 2.5$ for small objects and $q = 3$ for large objects.  
These power laws are connected by a transition region which
moves to larger radii with increasing $e_0$.  The fragmentation 
population thus extends to larger radii with {\it larger} $e_0$.  
In contrast, the merger population (the steep power law at 
larger radii) extends to smaller radii with {\it smaller} $e_0$.
This feature of the calculations can be tested directly with
observations (\cite{kl99})

\subsubsection{Models with Weaker Bodies}

The intrinsic strength $S_0$ of a planetesimal changes the growth
rate by setting the impact energy $Q_d$ needed to disrupt colliding 
bodies.  In our disruption model, $Q_d$ depends on the sum of $S_0$ 
and the gravitational binding energy of a colliding pair of 
planetesimals, $ 4 \pi G \rho_0 R_c^2/15$ (eq. [A11] and [A13]).  
The gravitational term is small compared to $S_0$  when the combined 
radius of the merged object is $R_c \lesssim 0.03 S_0^{1/2}$ km.  
The maximum radius of our initial size distribution, $r_0$ = 80 m, falls
below this limit for $S_0 \lesssim$ 25 erg g$^{-1}$. For collisions
between equal mass bodies, we can then derive a simple expression relating 
the minimum strength necessary for growth and the velocity dispersion, 

\begin{equation}
S_{0,min} \approx 6 \times 10^4 ~ {\rm erg~g^{-1}} \left ( \frac{V_{ij}}{100~{\rm m~s^{-1}}} \right ) ^2
\end{equation}

\noindent
At the start of our calculations, this result yields

\begin{equation}
S_{0,min} \approx 300~{\rm erg~g^{-1}} \left ( \frac{e_0}{10^{-3}} \right )^2 .
\end{equation}

\noindent
Bodies with $e_0 \lesssim$ 0.1 initially grow for the standard 
case with $S_0$ = $2 \times 10^6$ erg g$^{-1}$.  As these objects
grow to 1 km sizes, viscous stirring and dynamical friction increase 
the velocities of small objects.  Once the velocities reach the threshold 
set by Equation (1), catastrophic disruption produces debris that is 
lost from the calculation.  This process should limit the growth of 
the largest objects by reducing the reservoir of small bodies available 
for accretion.  We thus expect the maximum size of KBOs to depend on $S_0$.

To test these ideas in detail, we consider models with various $S_0$.
Calculations with $S_0$ = 10 to $2 \times 10^6$ erg g$^{-1}$ for $e_0 
= 10^{-4}$ and $S_0$ = $10^4$ to $2 \times 10^6$ erg g$^{-1}$ for $e_0 
= 10^{-3}$ allow velocity evolution to increase particle velocities up to 
the disruption threshold.  Models with $S_0 \lesssim 2 \times 10^6$ 
erg g$^{-1}$ and $e_0 \gtrsim 10^{-3}$ lose too much mass in the 
early stages to be of much practical value. Models with $S_0 \gtrsim 
2 \times 10^6$ erg g$^{-1}$ are identical to models with 
$S_0 = 2 \times 10^6$ erg g$^{-1}$.

Table 7 summarizes our results at 50 Myr and 100 Myr for $M_0 = 10~M_E$ 
and several different values of $\delta$.  For both values of $e_0$, 
stronger objects can grow to larger sizes at 100 Myr (see Figure 12).  
In each model, accretion and collisional damping lead to a short 
runaway growth phase that produces 100+ km objects with low 
velocities ($h_i \lesssim$ 0.1~m~s$^{-1}$), but leaves most of 
the initial mass in 0.1--1~km objects with much larger velocities
($h_i \sim$ 3--10~m~s$^{-1}$).  Dynamical friction and viscous 
stirring then increase the velocities of these small objects to 
the disruption threshold.  The timescale to reach this threshold 
increases with $S_0$; $r_{max}$ also increases with $S_0$ as indicated 
in Table 7 and Figure 12.  When all small bodies have been disrupted, 
the maximum radius is nearly independent of $\delta$:

\begin{equation}
{\rm log}~r_{max} \approx 2.45 + 0.22~{\rm log}~S_0 ~ 
\end{equation}

\noindent
for 10 erg g$^{-1}$ $\le S_0 \le 10^4$ erg g$^{-1}$ and
$e_0 = 10^{-4}$ and $10^{-3}$.  We did not run models with 
larger $S_0$ to the disruption threshold. Future calculations
will allow us to see whether $r_{max}$ reaches a threshold value at 
large $S_0$ or continues to increase as indicated by Equation (3). 

Although $r_{max}$ depends on $S_0$, both $r_5$ and the slope of 
the final size distribution at large radius are independent of $S_0$.  
The radius limit for $10^5$ KBOs has a small range, $r_5 \approx$ 
45--60 km, for any combination of $\delta$ and $S_0$ (Table 7).  
Figure 13 shows the evolution of the size distribution for $S_0$ 
= 10 erg g$^{-1}$.  As in Figure 8 and Figure 11, accretion produces 
a fragmentation tail with $q = 2.5$ at small radii and a steeper 
power law with $q = 3$ at large radii.  The $q = 3$ power law 
persists throughout the catastrophic disruption phase; the 
fragmentation tail evolves into a very steep power law with $q = 4$.
This behavior of the fragmentation tail occurs because larger objects
initially experience more disruptive collisions than do smaller objects.
Catastrophic disruption adds high velocity fragments from larger 
objects to the smaller mass bins; kinetic energy from this debris 
and viscous stirring gradually push smaller and smaller objects 
over the disruption threshold.  Eventually, the smallest objects in 
our grid, $r_i \approx$ 1 ~m, reach the disruption threshold and 
are slowly removed from the calculation.

\subsubsection{Models with Different Size Distributions}

The initial size distribution is one of the most uncertain input
parameters of our coagulation models.  The growth of 1--10 m or
larger bodies from interstellar dust grains is poorly understood.
Predicted sizes for conditions in the outer solar system range from 
$\lesssim$ 1 m up to several hundred km depending on details of both
microscopic and macroscopic physics (e.g., \cite{gol73}; \cite{tre90};
\cite{cz93a}, 1993b; \cite{bai94}; \cite{we97a}; \cite{bos97}; 
\cite{wur98}).  To test the sensitivity of our results to the 
initial conditions, we consider models with
(a) $q_0  = 4.5$, (b) $ q_0 = 1.5$, and 
(c) $N_C = {\rm Const}~\delta(r - r_0)$.
These models produce final size distributions that are indistinguishable 
from our standard model with $q_0 = 3$.  Both $r_{max}$ and $r_5$ are 
also independent of the initial size distribution.  The time to 
runaway growth and the timescale to produce one Pluto increase 
with the amount of material initially in the smallest objects, 
because collisional damping is more effective when there is a
large reservoir of small objects.  We derive 
$\tau_P$ = 30 Myr for $q_0 = 4.5$,
$\tau_P$ = 37 Myr for $q_0 = 3$,
$\tau_P$ = 42 Myr for $q_0 = 1.5$, and
$\tau_P$ = 49 Myr for $N_C = {\rm Const} ~ \delta(r - 80~m)$.

To test the importance of the mass spacing factor,
we recomputed these models for $\delta$ = 1.25.  Our results for
(a), (b), and (c) in the preceding paragraph are indistinguishable
from results with $\delta$ = 1.4, except that the timescale to 
produce Pluto decreases by $\sim$ 5\%.
The model with $\delta$ = 1.25 and our standard initial size distribution,
$q_0 = 3$, produced a single large body that ran away from the rest 
of the large bodies.  The final radius of this single object is 
$\sim$ 50\% larger than $r_{max}$ for other calculations with 
$M_0 = 10 M_E$.  Several tests indicate that forming a single 
large body is a stochastic process sensitive to $\delta$: 
small $\delta$ models produce such an object more often than 
large $\delta$ models.  We plan to investigate this possibility
further in future studies.

\subsubsection{Other Input Parameters}

To conclude this section, we briefly comment on the sensitivity 
of our calculations to other input parameters listed in Table 1.  
The results described above are insensitive to factor of two 
variations in the particle mass density $\rho_0$, the relative
gas velocity $\eta$, the minimum velocity for cratering $V_f$,
and the crushing energy $Q_c$.  We suspect that factor of ten 
variations in $\eta$, $V_f$, and $Q_c$ will also have no impact 
on the results. Larger variations in $\rho_0$ would probably change 
the variation of $r_{max}$ with $S_0$ in Table 7, but we have not 
investigated this possibility.  Increasing the fraction of kinetic
energy imparted to fragmentation debris $f_{KE}$ decreases the
time needed to produce 100+ km objects\footnote{In contrast, use
of the WS93 fragmentation algorithm increases the time needed to
produce 100+ km objects, although this increase is $\lesssim$
10\%--20\%}.  The growth time decreases by $\sim$ 10\% for 
$f_{KE}$ = 0.1 and $\sim$ 20\% for $f_{KE}$ = 0.2.  Variations in 
the input $f_{KE}$ do not change the final size distribution or 
the number of KBOs as a function of $M_0$.  This parameter thus 
has less impact on the results than either $e_0$ or $S_0$.  

\subsection{Limitations of the Models}

We summarized the major uncertainties of our planetesimal calculations
in KL98 and will repeat important points for the current models here.
Our choice of a single accumulation zone does not allow us to follow
the evolution in semimajor axis of a planetesimal swarm (\cite{spa91};
\cite{we97b}). Although multi-zone calculations are important for 
understanding migration and other long-term aspects of planetesimal
evolution, single-annulus models are a reasonable first approximation 
for the early evolution of KBOs.  The growth of large nearby bodies, 
such as those that will merge to form Neptune, should modify the 
velocity evolution of KBOs once these bodies reach sizes much 
larger than 1000 km (see \cite{mor97}).  For most solar nebulae, 
these long-distance interactions should remain small until Pluto 
forms beyond 30 AU (KL98; see also \cite{laz94}; \cite{roq94}; 
\cite{mor97}; \cite{war98}).  Calculations now underway will 
test this assertion in more detail.

Our relatively coarse mass grid probably overestimates the timescale
to produce KBOs and Pluto by $\sim$ 5\%--10\% (KL98).  
The delay in runaway growth relative to a $\delta = 1.25$ model 
is 3\%--5\% for $\delta$ = 1.4 and 5\%--10\% for $\delta$ = 2.0.  
Although these delays are small compared to the overall uncertainties 
in our algorithms, calculations with $\delta \gtrsim$ 2 rarely 
produce one or more isolated bodies that grow much faster than 
smaller bodies.  The better mass resolution of $\delta \lesssim$ 1.4 
calculations allows more mass batches to satisfy the isolation 
condition, defined in the Appendix, which leads to more accurate 
calculations of the cumulative mass distribution and the velocity 
evolution of the lowest mass objects.

The lack of a rigorous treatment of gas dynamics probably has little
impact on our results.  Gas drag removes $\lesssim$ 1\% of the initial
mass from these models.  Gas accretion by large bodies is also 
insignificant.  The minimum radius needed to capture gas from the 
disk is $r_i \sim$ 1000--2000 km for typical temperatures of 
50--100 K at 30--50 AU (\cite{bec90}; \cite{ost94}).  
Our models reach this limit on timescales exceeding the lifetime of 
the gaseous disk, so we expect little gas accretion by Kuiper Belt
bodies.

Our calculations probably overestimate the amount of mass lost to 
dust. At both 1 AU and 35 AU, losses from catastrophic fragmentation, 
cratering, and gas drag grow with increasing $\delta$.  
In large $\delta$ models, small delays in the formation of runaway 
bodies allow dynamical friction and viscous stirring extra time to 
increase the velocity dispersions, and hence mass loss, of the smallest 
objects.  These effects are probably small, $\sim$ 10\%--20\%, for 
most of our calculations.

Our choice of the initial size distribution has little impact 
on our results.  Models with $q_0 = 1.5$ and $ q_0 = 4.5$ produce 
final size distributions very similar to those for calculations with 
$q_0 = 3$.  The timescale to produce 1000 km objects lengthens 
as $q_0$ decreases.  The `equilibrium' size distribution with
$q = 3$ is similar to the observed size distribution of interstellar 
dust grains, $q \approx$ 2--3 (e.g, \cite{kmh94}; \cite{li97} and 
references therein).  Similar size distributions are derived from 
calculations of the growth of very small particles using measured 
sticking efficiencies (e.g., \cite{wur98}).

Aside from the timescale to produce 1000 km objects and the amount 
of mass lost to gas drag and fragmentation, the initial eccentricity 
distribution probably also has little impact on our conclusions.  
The number of 50 km radius KBOs is not sensitive to $e_0$;
$r_{95\%}$ increases with $e_0$ only for $e_0 \gtrsim 10^{-2}$ 
(Table 6).  The slope of the final size distribution is also 
insensitive to $e_0$.  

We suspect that initial eccentricities outside the range considered 
here are unrealistic.  Viscous stirring and gas drag appear to set a 
lower limit on the eccentricity of small objects, $e \sim 10^{-5}$.
Models with $e_0 \lesssim 10^{-4}$ should thus closely resemble 
$e_0$ = $10^{-4}$ models.  For $e \gtrsim 10^{-2}$, collisions 
lead to substantial fragmentation and mass loss from the numerical 
grid. Circularization is then less effective and growth is very slow 
(see Table 6). These calculations probably poorly approximate reality:
dynamical friction between 1--100 m objects and smaller dust particles 
not included in our grid should reduce the velocities of 1--100 m
objects on very short timescales\footnote{This problem does not occur
in models with small $e_0$, because the mass loss is $\lesssim$ 
1\% of the initial mass.}.  We expect that these calculations would
then more closely resemble models with smaller $e_0$, if dust 
particles can grow back into 1 m bodies.

The final limitation of our model is the fragmentation algorithm.
We adopt an energy-scaling fragmentation law, because other types
of models have not been developed and tested for the low velocity
conditions appropriate in the Kuiper Belt. Other scaling models seem 
to yield better results for main belt asteroids than do energy-scaling
models, but these models assume that the collision can be approximated
as a point-like impact on a large body (\cite{dav94}; \cite{dur98};
\cite{rya98}).  This assumption is quite good for the high speed 
impacts, $\gtrsim$ 100--500 m s$^{-1}$, of strong, dense objects 
in the inner solar system.  Point-like impacts are probably rare for 
the lower velocity collisions of weaker, low density objects like KBOs. 
Our consideration of KBOs with a large range of intrinsic strengths
suggests that an improved fragmentation model would not change our
results significantly.

\section{DISCUSSION AND SUMMARY}

In KL98 and this paper, we have developed a time-dependent planetesimal
evolution code to calculate the formation of KBOs in a single annulus
outside the orbit of Neptune.  The computer program includes coagulation
with realistic cross-sections, energy-scaling algorithms to treat
cratering and disruption, and velocity evolution using the statistical
expressions of Hornung \etal (1985). Our numerical solutions match 
standard analytic test cases and generally reproduce the results 
of other accretion and collision calculations (e.g., WS93; \cite{dav97}).

Our calculations demonstrate that plausible models can satisfy current
observations of the Kuiper Belt.  Several Plutos and $\sim 10^5$ 
50 km radius KBOs form in Minimum Mass Solar Nebulae with 
$e_0 \approx 10^{-3}$ on timescales of 20--40 Myr.  Growth is more 
rapid in more massive nebulae and in planetesimal swarms with 
smaller initial velocities. The formation time is less sensitive 
to the initial size distribution, the intrinsic strength of KBOs, 
and other input parameters listed in Table 1.

Each Kuiper Belt model yields a cumulative size distribution with 
two main features.  Objects with $r_i \lesssim$ 0.1 km follow 
$N_C \propto r^{-q}$ with $q = 2.5$, as expected for collision 
fragments (\cite{doh69}; \cite{wil94}). Larger objects with 
$r_i \gtrsim$ 1--10 km follow a $q \approx 3$ power law over several
orders of magnitude in radius.  These slopes do not depend on 
$M_0$, $e_0$, $S_0$, $f_{KE}$, and $q_0$, among other input 
parameters.  Kenyon \& Luu (1999) compare these results with 
observations.  Here, we note that the $q \approx 3$ power law 
for large bodies is identical to the observed slope, $3 \pm 0.5$ 
(\cite{jew98}; \cite{luu98}; but see also \cite{gla98}).  

Fragmentation and velocity evolution are important components in the 
formation of present day KBOs.  Fragmentation produces a large reservoir 
of small bodies that damp the velocity dispersions of the large objects 
through dynamical friction.  These processes allow a short runaway growth 
phase where 1 km objects grow into 100 km objects.  Continued fragmentation
and velocity evolution damp runaway growth by increasing the velocity 
dispersions of small objects.  This evolution leaves $\sim$ 1\%--2\% 
of the initial mass in 50 km radius KBOs.  The remaining mass is in 
0.1--10~km radius objects.  Fragmentation will gradually erode these 
smaller objects into dust grains that are removed from the Kuiper Belt 
on short timescales, $\sim 10^7$ yr (see \cite{bac93}; \cite{bac95}).  
Thus, 50 km radius KBOs comprise a small fraction of the original 
Kuiper Belt.

Fragmentation also limits the size of the largest object in the Kuiper 
Belt.  The maximum radius ranges from $r_{max} \approx$ 450 km for 
$S_0$ = 10 erg g$^{-1}$ to $r_{max} \gtrsim$ 3000 km for 
$S_0$ = $2 \times 10^6$ erg g$^{-1}$.  Pluto formation sets 
a lower limit on the tensile strength of KBOs, $S_0 \ge 300$ erg g$^{-1}$.

These results suggest a refinement of our picture for KBO formation
in the outer solar system.  In KL98, we speculated that velocity
perturbations due to the growth and outward migration of Neptune 
would limit the growth of KBOs at radii $\lesssim$ 1000 km.  Although 
this hypothesis is plausible (see, for example, \cite{mal93}; \cite{mor97}), 
our current models demonstrate that 50--1000 km radius KBOs form naturally 
at $\sim$ 35 AU on timescales, $\tau_P \sim$ 10--100 Myr, comparable 
to the Neptune formation time.  Although a few objects can reach 
2000+ km radii on timescales of 2--3 $\tau_P$, nearly all of the 
initial mass beyond 30 AU remains in small, 1 km radius objects that 
can be depleted by collisional disruption (\cite{dav97}) or gravitational 
sculpting (\cite{hol93}) or both on timescales exceeding 100 Myr.  
This evolution can account for the observation of 50+ km KBOs in a 
currently small mass Kuiper Belt without intervention by Neptune.

Finally, our new results further support the notion that KBOs will form 
in other solar systems.  The dusty circumstellar disks detected in many 
pre--main-sequence stars suggest masses of 1--100 $M_E$ at distances
of 30--100 AU (e.g., \cite{sar93}; \cite{bec96}; see also \cite{clo97}; 
\cite{hoh97}; \cite{lay97}; \cite{ake98}; \cite{sta98}).  KBOs with
50+ km radii can grow in this material as the central stars contract 
to the main-sequence if the disks are not too turbulent (see 
\cite{cz93a}, 1993b).  Smaller, 1--10 km radius KBOs probably form 
in less massive pre--main-sequence disks.  Our results also
indicate that the growth of 100--1000 km radius KBOs is accompanied 
by substantial dust production, $\sim$ 0.1--1 $M_E$, in models with
$M_0 \sim$ 10--100 $M_E$.  This dust could be responsible for the 
ringlike structures observed in pre--main-sequence stars such as
GG Tau (\cite{rod96}) and older main sequence stars such as 
$\epsilon$ Eri (\cite{gre98}) and HR 4796 (\cite{jay98}; \cite{koe98}).  
The less massive disks in $\alpha$ Lyr, $\alpha$ PsA, and 
$\beta$ Pic may also harbor KBOs if the dust masses are 
reasonably close to the `maximum' masses inferred for these 
systems (\cite{bac93}). In all of these stars, the dynamics and 
mass distribution of dust may well provide useful constraints on 
the properties of presumed KBOs.  We hope to explore this possibility
in future studies.

\vskip 4ex

We thank B. Bromley for making it possible to run our code on the 
JPL Cray T3D `Cosmos' and the HP Exemplar `Neptune' and for a 
generous allotment of computer time through funding from the 
NASA Offices of Mission to Planet Earth, Aeronautics, and Space Science.
Comments from A. Cameron, F. Franklin, M. Geller, M. Holman, 
S. Starrfield, and J. Wood greatly improved our presentation.  
We acknowledge G. Stewart for clarifying details of the WS93 calculations.  

\appendix

\section{APPENDIX}

\subsection{Overview}

As described in KL98, our accretion model assumes that planetesimals 
are a statistical ensemble of masses in a cylindrical annulus of width
$\Delta a$ and height $H$ centered at a radius $a$ from the Sun.
The particles have horizontal $h_i(t)$ and vertical $v_i(t)$
velocity dispersions relative to an orbit with mean Keplerian velocity 
$V_K$ (see \cite{lis93}).  We approximate the continuous distribution 
of particle masses with discrete batches having an integral number of 
particles $n_i(t)$ and total masses $M_i(t)$ (\cite{ws93}).  
The average mass of a batch, $m_i(t)$ = $M_i(t) / n_i(t)$, evolves 
with time as collisions add and remove bodies from the batch.  

To compute the evolution of particle numbers and velocities in 
KL98, we solved the coagulation equation and a set of velocity
evolution equations for all mass bins $k$ during a time step
$\delta t$.  We assumed that bodies merge but do not fragment
during collisions.  We conserved kinetic energy in each collision
and adopted a kinetic approximation to calculate velocity changes
due to gas drag, dynamical friction, and viscous stirring.  Our
explicit algorithm for solving these equations reproduced standard 
tests and other published calculations.

In this paper, we consider collisions that produce mergers and debris.
The coagulation equation is then:

\begin{equation}
\delta n_k = \delta t \left [ \epsilon_{ij} A_{ij} n_i n_j ~ - ~ n_k A_{ik} n_i \right ] ~ + ~ \delta n_{k,f}~ - ~ \delta n_{k,gd}
\end{equation}

\begin{equation}
\delta M_k = \delta t \left [ \epsilon_{ij} A_{ij} n_i n_j m_k ~ - ~ n_k A_{ik} n_i m_k \right ] ~ + ~ m_k \delta n_{k,f} - ~ m_k \delta n_{k,gd}
\end{equation}

\noindent
where $A_{ij}$ is the cross-section,
$\epsilon_{ij} = 1/2$ for $i = j$ and $\epsilon_{ij} = 1$ for $i \ne j$.
The four terms in A1--A2 represent (i) mergers of $m_i$ and $m_j$ into 
a body of mass $m_k = m_i + m_j - m_{e,ij}$, (ii) loss of $m_k$ through 
mergers with other bodies, (iii) addition of $m_k$ from debris produced
by the collisions of other bodies, and (iv) loss of $m_k$ by gas drag.
We consider below the mass lost to small fragments $m_{e,ij}$.
The second term in A1--A2 includes the possibility that a collision 
can produce debris but no merger (rebounds).  As in KL98, we 
calculate the ``gravitational range'' of the largest bodies -- 
$R_{g,i} = K_1 a R_{H,ii_{mid}} + 2 a e_i$ (\cite{ws93}) --
where $K_1 = 2 \sqrt{3}$ and $R_{H,ij} = [(m_i + m_j)/3 \msun]^{1/3}$
is the mutual Hill radius.  As in WS93, the isolated bodies are the 
$N$ largest bodies that satisfy the summation, 
$ \sum_{i_{min}}^{i_{max}} ~ n_i R_{g,i} \ge \Delta a$.
These isolated bodies cannot collide with one another but can 
collide with other lower mass bodies.

As in KL98, we solve the complete set of evolution equations, 
A1--A2 above and A7--A8 from KL98, using an explicit method that 
limits the time step automatically to prevent large changes in the 
dynamical variables.  Section 2 of the main text compares calculations 
at 1 AU with results from WS93.  In the rest of this Appendix, we describe
fragmentation algorithms (A.2) and updates to our velocity evolution
algorithm (A.3).

\subsection{Fragmentation}

Algorithms for collision outcomes rely on comparisons between the 
kinetic energy of the two colliding planetesimals $Q_{c,ij}$ and 
the binding energy of the merged planetesimal $Q_{b,ij}$. The 
binding energy usually consists of an intrinsic tensile strength and the
gravitational binding energy, $Q_{b,ij} = S_0 + Q_{g,ij}$. This `energy 
scaling' approach is sometimes replaced by other scaling laws to 
model the structure of the colliding bodies more accurately.  For example,
Housen and collaborators (see \cite{hou90}; \cite{hou91}; \cite{hls93}, 1994)
describe a strain-rate scaling model to express the collision energy 
needed to disrupt a body in terms of its size and impact velocity.
Davis \etal (1994) showed that both types of model can match observations 
of the asteroid belt in our solar system (see also \cite{far82};
\cite{dav89}; \cite{wil94}; \cite{mar97}).

In this paper, we adopt an energy scaling model.  Energy scaling has 
two main advantages for calculating the fragmentation of Kuiper Belt
objects.  The wide use of this approach allows us to compare results 
with many previous calculations.  Other models also seem inappropriate 
for the low velocity collisions anticipated in the Kuiper Belt.  
These models have been scaled for conditions in the inner 
Solar System, where collision velocities exceed 100 m s$^{-1}$ 
(see \cite{dav94}).  The approximations made for these collision
velocities fail when the impact velocity is smaller than 100--1000 
m s$^{-1}$ (e.g., \cite{hou90}, p. 239).  We expect velocities of 
$\lesssim$ 10 m s$^{-1}$ for most collisions in the Kuiper Belt 
(KL98), and thus cannot apply the Housen \etal (1990) and 
other sophisticated models in our calculations.

The output of any fragmentation algorithm is the mass of the
merger remnant $m_k$ and the mass distribution of the fragments.
We assume $m_k = m_i + m_j - m_{e,ij}$, where $m_{e,ij}$ is the
mass of fragments {\it ejected} from the merged planetesimal.
We consider two approaches to compute $m_{e,ij}$.
In the first case, we follow WS93 and set the impact velocity as

\begin{equation}
V_{I,ij}^2 = V_{ij}^2 + V_{e,ij}^2
\end{equation}

\noindent
where $V_{ij}$ is the relative velocity of the two colliding bodies 
(equation A12 of KL98) and $V_{e,ij}^2$ = $2 G (m_i + m_j)/(r_i + r_j)$ 
is the mutual escape velocity.  The center of mass fragmentation
energy $Q_{f,ij}$ and the gravitational binding energy $Q_{g,ij}$
per unit mass are

\begin{equation}
Q_{f,ij} = (K_2/2)~m_i~m_j~V_{I,ij}^2 / (m_i + m_j)^2
\end{equation}

\noindent
and

\begin{equation}
Q_{g,ij} = 0.6 ~ G ~ (m_i + m_j) / R_c
\end{equation}

\noindent
where $R_c$ is the spherical radius of the combined body with mass 
$m_i + m_j$ and $K_2$ = 0.5 is a constant.  We further define the 
fragmentation energy $E_{f,ij} = (m_i + m_j) Q_{f,ij} $.

WS93 and most other studies assume that collisions produce debris when 
(i) the impact velocity exceeds a threshold velocity, $V_{I,ij} > V_f$,
and (ii) the amount of ejected mass exceeds a threshold value,
$m_{e,ij} > 10^{-8} (m_i + m_j)$.  The mass ejected in the collision
is derived from simple energy considerations.  The collision 
{\it disrupts} the merged object if the collision energy exceeds
the binding energy, $Q_{f,ij} > Q_{g,ij} + S_0$, where $S_0$ is 
the impact strength.  In most of their calculations, WS93 further 
require that the mass fragmented by the impact exceed half of 
the mass of the merged object, $m_{f,ij} > 0.5 (m_i + m_j)$,
where the fragmented mass is (\cite{gre78}, 84):

\begin{equation}
m_{f,ij} = Q_{f,ij} / Q_c ~ ,
\end{equation}

\noindent
and $Q_c$ is the crushing energy. If both conditions for 
``catastrophic disruption'' are met, the collision produces 
a large fragment with mass (Fujiwara \etal 1977; see also
\cite{fuj80}, \cite{hou90}): 

\begin{equation}
m_{L,ij} = 0.5~(mi + mj)~(Q_{m,ij}/(Q_{g,ij}~+~Q_b))^{-1.24} ~ ,
\end{equation}

\noindent
and numerous smaller fragments.  In all cases, $m_{L,ij} < 0.5~(m_i + m_j)$.

If the conditions for catastrophic disruption are not met, the collision 
produces a large body with $m_k \approx m_i + m_j$ and numerous small 
fragments with total mass $m_{f,ij}$ (A6).  WS93 assume that the amount 
of the fragmented mass that escapes the merged body is

\begin{equation}
m_{e,ij} = K_3 m_{f,ij} V_{e,ij}^{-2.25}
\end{equation}

\noindent
where $K_3 = 3~\times~10^6$ (cm s$^{-1})^{2.25}$ is a constant
(see also \cite{gau63}; \cite{gre78}).  The mass of the largest 
fragment in this case is

\begin{equation}
m_{L,ij} = 0.2 ~ m_{e,ij} ~ .
\end{equation}

\noindent
In WS93, the mass distribution of the debris depends on the method
of fragmentation.  We simplify this procedure and adopt a single
power law distribution for both methods:

\begin{equation}
\delta n_k = C_k ( m_1^{-b} - m_2^{-b})
\end{equation}

\noindent
where $b = (1 + m_{L,ij}/m_{e,ij})^{-1}$,
$C_k = m_{L,ij}^b$,
$m_1 = (m_{k} m_{k-1})^{1/2}$, and
$m_2 = (m_{k} m_{k+1})^{1/2}$ (\cite{gre78}).
Tests show that this distribution produces results similar 
to those of WS93 when we assume that the velocity of the
escaping fragments equals the relative velocity of the 
colliding planetesimals, $V_{ij}$ (see section 2).

As a comparison to the WS93 fragmentation algorithm, we consider
the Davis \etal (1985; \cite{gre78}, 1984; \cite{dav94}) approach. 
We adopt their energy scaling formula and write the strength 
(in erg g$^{-1}$) of a planetesimal as:

\begin{equation}
S = S_0 + \frac{4 \pi K_4 G \rho_0 R_c^2}{15} ~ ,
\end{equation}

\noindent
where $K_4$ is a constant.  We adopt $K_4$ = 1 in the absence of any 
useful estimates for Kuiper Belt objects; Davis \etal (1985, 1994) 
consider $K_4$ = 1--100 for collisional evolution in the asteroid
belt (see also \cite{hou90}). Davis \etal assume that the ejecta 
receive a fixed fraction $f_{KE} Q_{f,ij}$ of the impact energy
and have a power law mass-velocity distribution,

\begin{equation}
f(>v) \propto (v/v_c)^{-\alpha_V} ~ ,
\end{equation}

\noindent
where $v_c$ is a reference velocity. The impact energy needed to 
disrupt the planetesimal and accelerate 50\% of the fragments to
escape velocity follows from integration of (A12) and energy
conservation (\cite{dav85}):

\begin{equation}
Q_d = S ~ \left ( \frac{\alpha_V~0.5^{1~+~2/\alpha_V}}{f_{KE} (\alpha_V - 2)} \right ) ~ .
\end{equation}

\noindent

Davis \etal assume that collisions produce debris
when the impact velocity exceeds a threshold, $V_{I,ij} > V_f$.
We also adopt the WS93 threshold for the ejected mass,
$m_{e,ij} > 10^{-8} (m_i + m_j)$. 
The collision disrupts the colliding bodies if $Q_{m,ij} > Q_d$.
If this condition is met, the total mass lost is:

\begin{equation}
m_{e,ij} = 0.5~(m_i + m_j)~(Q_{m,ij}/Q_d)^{0.5\alpha_V} ~ ,
\end{equation}

\noindent{while the mass of the largest fragment is the smaller of}

\begin{equation}
m_{L,ij} = \left\{ \begin{array}{l l}
           0.5~(m_i + m_j)~(Q_{m,ij}/Q_d)^{1-\alpha_V} & \\
           0.2 ~ m_{e,ij} & \\
         \end{array}
         \right .
\end{equation}

\noindent
If the collision does not disrupt the merged planetesimal,
the mass in fragments is simply $m_{f,ij}$ from (A6). To estimate
the fraction of this mass that escapes, we follow the Davis \etal (1985)
derivation of $Q_d$ and integrate the mass distribution over velocity:

\begin{equation}
m_{e,ij} = \left ( \frac{f_{KE}~(\alpha_V - 2) m_j V_{I,ij}^2}
                        {\alpha_V~m_{f,ij}~V_{e,ij}^2} \right )^{0.5\alpha_V}
\end{equation}

\noindent
The mass of the largest fragment is then $m_{L,ij} = 0.2 ~ m_{e,ij}$
as before.  We calculate the mass distribution of the fragments as 
in (A10) and assume that all fragments have the same kinetic energy 
per unit mass.

Unlike WS93, Davis \etal allow for `rebounds,' where the colliding
bodies do not merge into a single body. We follow Barge \& Pellat
(1992) and define the rebound velocity as

\begin{equation}
V_{reb, ij}^2 = \frac{2(1 - c_R^2)}{c_R^2} V_{e,ij}^2
\end{equation}

\noindent
where $c_R$ is the coefficient of restitution. In most applications,
$c_R$ takes on separate values for collisions with impact velocities
above and below the fragmentation threshold:

\begin{equation}
c_{R} = \left\{ \begin{array}{l l l}
         c_1 & \hspace{5mm} & V_{I,ij} < V_f \\
         c_2 & \hspace{5mm} & V_{I,ij} \ge V_f \\
         \end{array}
         \right . 
\end{equation}

\noindent
With these definitions, collisions with $V_{I,ij} < V_{reb, ij}$
produce mergers and debris; collisions with 
$V_{I,ij} > V_{reb, ij}$ produce debris but no merger.  We assume
that rebounds conserve energy and that the velocity difference
between bodies after the collision is 

\begin{equation}
(v_i - v_j)_{after} = c_R (v_i - v_j)_{before}.
\end{equation}

\noindent
Tests of the coagulation code indicate that this prescription 
yields results almost identical to a prescription where we 
assume that the particle velocities are not changed by the
rebound.

In addition to the tests described in section 2, we repeat the
collisional evolution of KBOs described in Davis \& Farinella (1997).
With suitable modifications to our algorithms for the collision
cross-section and velocity evolution, we reproduce their results
to $\sim$ 10\%--20\%.

\subsection{Velocity evolution}

Our procedure for treating the evolution of the horizontal $h_i$
and vertical $v_i$ components of the velocity dispersion follows 
the kinetic formalism developed by Hornung \etal (1985) and
Barge \& Pellat (1990, 1991, 1992).  In KL98, we adopted the
WS93 expressions for dynamical friction and viscous stirring due
to long range encounters and reformulated the Barge \& Pellat (1990, 
1991) expressions for inelastic collisions in terms of $e$ and $i$ 
(see equations A19--A22 of KL98).  We verified that our algorithm
for computing the velocity evolution achieved the equilibrium
value for the ratio of the vertical to the horizontal velocity
$\beta = i/e$ = 0.6 (\cite{hor85}) and generally reproduced the
velocity evolution of the WS93 Earth calculation.

In this paper, we make two modifications to our treatment of 
the velocity evolution and clarify our treatment of velocity
evolution at low velocities.  We assume that rebound collisions
do not contribute to dynamical friction and viscous stirring for
inelastic collisions (equations A21 and A22 of KL98). With
this assumption, the orbits of low mass objects do not circularize 
when their velocity dispersion exceeds the rebound velocity (equation
A17 above).  This behavior slows down the growth time of massive
objects, because the collision cross-section remains in the high
velocity limit.

Although the kinetic prescription of Hornung \etal (1985) accurately
represents dynamical evolution at high particle velocities,
it fails as the particle velocity approaches the Hill velocity
$v_{H,ij} = \Omega R_{H,ij}$ (\cite{ida90}; \cite{bar91}; WS93).
Ida (1990) used $N$-body calculations to derive the appropriate
behavior at low particle velocities, $V_{ij} < $ 2--5 $v_{H,ij}$.
We generally reproduce Ida's (1990) results if we adopt

\begin{equation}
\left ( \frac{de_{vs,i}^2}{dt} \right )_{lv} = \left ( \frac{e_i^2~+~e_j^2}{e_{ij,lv}} \right ) ~ \left ( \frac{de_{vs,i}^2}{dt} \right )_{hv} ~~,~~
\left ( \frac{di_{vs,i}^2}{dt} \right )_{lv} = \left ( \frac{e_i^2~+~e_j^2}{e_{ij,lv}} \right )^2 ~ \left ( \frac{di_{vs,i}^2}{dt} \right )_{hv}
\end{equation}

\noindent
for viscous stirring and

\begin{equation}
\left ( \frac{de_{df,i}^2}{dt} \right )_{lv} = \left ( \frac{e_i^2~+~e_j^2}{e_{ij,lv}} \right )^2 ~ \left ( \frac{de_{df,i}^2}{dt} \right )_{hv} ~~,~~
\left ( \frac{di_{df,i}^2}{dt} \right )_{lv} = \left ( \frac{e_i^2~+~e_j^2}{e_{ij,lv}} \right )^2 ~ \left ( \frac{di_{df,i}^2}{dt} \right )_{hv}
\end{equation}

\noindent 
for dynamical friction (see also \cite{bar91}, \cite{ws93}).  
In both equations, the `hv' subscript indicates the appropriate 
high velocity expression from KL98 (equations A19 and A20) 
and $e_{ij,lv}$ is the value of $e_i^2~+~e_j^2$ at $V_{ij} = V_{lv}~v_{H,ij}$.  
These equations yield constant timescales for dynamical friction
and the inclination component of viscous stirring (\cite{ida90};
Figure 10).  The timescale for the eccentricity component of 
viscous stirring varies as $e^2$ at low velocity (\cite{ida90}). 

We describe the sensitivity of the velocity evolution to the choice of
$V_{lv}$ in Sec. 2.  Any $V_{lv} \gtrsim $ 2 slows down velocity
evolution significantly, because the most massive particles in our 
test calculations have $V_{ij} \lesssim$ 0.3--0.5 $v_{H,ij}$.  The 
velocity changes are then $\sim$ 25 times smaller for viscous stirring
and $\sim$ 500 times smaller for dynamical friction (see also 
\cite{ida90}, \cite{bar91}, \cite{ws93}).  Our results are not
very sensitive to $V_{lv}$ for $2 \lesssim V_{lv} \lesssim 5$:
the velocity evolution slows down gradually as $V_{lv}$ increases
from 2 to 5 (see Sec. 2).  We adopt the middle of Ida's (1990) range,
$V_{lv}$ = 3.5, in our Kuiper Belt calculations. For comparison,
WS93 adopt $V_{lv}$ = 2; \cite{bar91} adopt
$V_{lv}$ = 0.8 for viscous stirring and
$V_{lv}$ = 2.6 for dynamical friction.

In many of our Kuiper Belt calculations, the velocity of the largest
bodies approaches zero due to collisional damping, dynamical friction, 
and fragmentation. 
To avoid divergences in the stirring rates from long-range encounters,
we adopt a minimum horizontal velocity of $10^{-3}$ m s$^{-1}$ and
a minimum vertical velocity of $5.3 \times 10^{-4}$ m s$^{-1}$.
This `floor' to the velocity evolution maintains the appropriate
$\beta_{ij}$ for collisions with $V_{ij} > V_{lv}~v_{H,ij}$ and
allows larger time steps for the brief interval when growth is rapid 
and the viscous stirring rate is small.  Tests show that the growth 
rate of the largest bodies does not depend on the values for these 
limits.  Lower values for the `floor' result in somewhat higher 
velocities for the small mass objects, but these changes are small.  
Once viscous stirring begins to dominate collisional cooling, the 
velocities of all bodies increase above our `floor' values.

Finally, to reduce the computation time of our velocity evolution algorithm,
we approximate the $I$, $J$, and $K$ integrals in equations A19--A22 of
KL98 with polynomial expressions derived using 
{\it Mathematica}\footnote{Mathematica $v3.0.2$ \copyright 1988-1997
Wolfram Research, Inc.}
Our results match the integral expressions to better than 1 part 
in $10^3$ for $\beta_{ij}$ = 0--1. 
Expressions ready for use in a computer calculation appear below.

\begin{eqnarray*}
I_r & = & 14.1439~+~0.0675783/\beta_{ij}~+~\beta_{ij}~*~(3.86566~+~\beta_{ij}~*~({-6.5623} \\
    &   & ~~~~~~~~~~+~\beta_{ij}~*~(12.8084~+~\beta_{ij}~*~({-9.93223}~+~2.94993~*~\beta_{ij})))) ~~~~~~~ {\rm (A21)} \\
\end{eqnarray*} 
\begin{eqnarray*}
I{_\theta} & = & 2.32693~+~0.0337724/\beta_{ij}~+~~\beta_{ij}~*~(1.80115~+~\beta_{ij}~*~({-3.3792} \\
    &   & ~~~~~~~~~~+~\beta_{ij}~*~(5.78079~+~\beta_{ij}~*~({-4.43822}~+~1.32431*~\beta_{ij})))) ~~~~~~~~ {\rm (A22)} \\
\end{eqnarray*} 
\begin{eqnarray*}
I_z & = & {-0.00137043}~+~\beta_{ij}*(0.121347~+~\beta_{ij}~*~(7.27632~+~\beta_{ij}~*~(6.32281 \\
    &   & ~~~~~~~~~~~+~\beta_{ij}~*~(4.98011~-~1.36123~*~\beta_{ij}))))  ~~~~~~~~~~~~~~~~~~~~~~~~~~~~~~~~~ {\rm (A23)} \\
\end{eqnarray*} 
\begin{eqnarray*}
J_r & = & {-4.21891}~+~\beta_{ij}~*~({-0.762162}~+~\beta_{ij}~*~(37.2078~+~\beta_{ij}~*~({-97.5891} \\
    &   &  ~~~~~~~~~~~~~~+~\beta_{ij}~*~(120.258~+~\beta_{ij}~*~({-72.0619}~+~\beta_{ij}~*~(14.8515 \\
    &   &  ~~~~~~~~~~~~~~+~1.57744*\beta_{ij})))))) ~~~~~~~~~~~~~~~~~~~~~~~~~~~~~~~~~~~~~~~~~~~~ {\rm (A24)} \\
\end{eqnarray*}
\begin{eqnarray*}
J_{\theta} & = & {-1.90771}~+~\beta_{ij}~*~(13.8618~+~\beta_{ij}~*~({-30.8032}~+~\beta_{ij}~*~(47.3479 \\
    &   &  ~~~~~~~~~~~~+~\beta_{ij}~*~({-49.4466}~+~\beta_{ij}~*~(32.4446~+~\beta_{ij}~*~({-11.7370} \\
    &   &  ~~~~~~~~~~~~+~1.72263~*~\beta_{ij})))))) ~~~~~~~~~~~~~~~~~~~~~~~~~~~~~~~~~~~~~~~~~~~ {\rm (A25)} \\
\end{eqnarray*} 
\begin{eqnarray*}
J_z & = & 6.12664~+~\beta_{ij}~*~({-13.1003}~+~\beta_{ij}~*~({-6.3946}~+~\beta_{ij}~*~(50.1757 \\
    &   &  ~~~~~~~~~~~+~\beta_{ij}~*~({-70.6068}~+~\beta_{ij}~*~(39.2953~+~\beta_{ij}~*~({-2.86619} \\
    &   &  ~~~~~~~~~~~-~3.37452~*~\beta_{ij})))))) ~~~~~~~~~~~~~~~~~~~~~~~~~~~~~~~~~~~~~~~~~~~ {\rm (A26)} \\
\end{eqnarray*} 
\begin{eqnarray*}
K_r & = & 3.16452~+~\beta_{ij}~*~(0.0149191~+~\beta_{ij}~*~({-1.73044}~+~\beta_{ij}~*~(1.55516 \\
    &   & ~~~~~~~~~~+~\beta_{ij}~*~(1.08438~+~\beta_{ij}~*~({-3.55476}~+~\beta_{ij}~*~(2.91763 \\
    &   & ~~~~~~~~~~-~0.841079~*~\beta_{ij})))))) ~~~~~~~~~~~~~~~~~~~~~~~~~~~~~~~~~~~~~~~~~~~~ {\rm (A27)} \\
\end{eqnarray*} 
\begin{eqnarray*}
K_\theta & = & 2.96882~+~0.135176/\beta_{ij}~+~\beta_{ij}~*~(7.13193~+~\beta_{ij}~*~({-24.1086} \\
    &   & ~~~~~~~~~~~~+~\beta_{ij}~*~(39.3287~+~\beta_{ij}~*~({-35.5738}~+~\beta_{ij}~*~(15.0527 \\
    &   & ~~~~~~~~~~~~+~\beta_{ij}~*~({-0.327693}~-~1.25529*~\beta_{ij}))))))  ~~~~~~~~~~~~~~ {\rm (A28)} \\
\end{eqnarray*} 
\begin{eqnarray*}
K_z & = & 6.13154~+~0.13519/\beta_{ij}~+~\beta_{ij}~*~({-6.70136}~+~\beta_{ij}~*~(4.49227 \\
    &   & ~~~~~~~~~~~~+~\beta_{ij}~*~({-6.53516}~+~\beta_{ij}~*~(19.317~+~\beta_{ij}~*~({-29.5368} \\
    &   & ~~~~~~~~~~~~+~\beta_{ij}~*~(21.0967~-~5.78973*~\beta_{ij})))))) ~~~~~~~~~~~~~~~~~~ {\rm (A29)} \\
\end{eqnarray*} 

\vfill
\eject

\epsffile{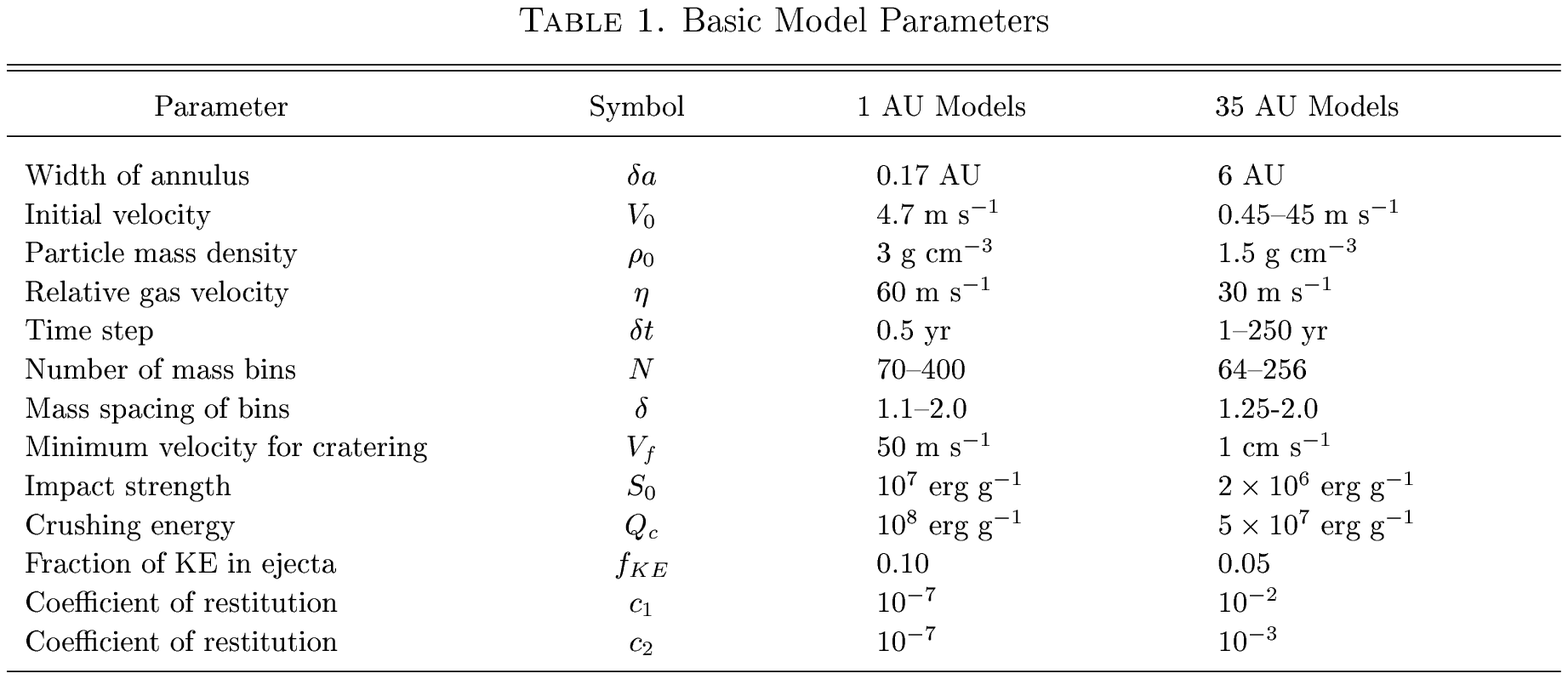}

\epsffile{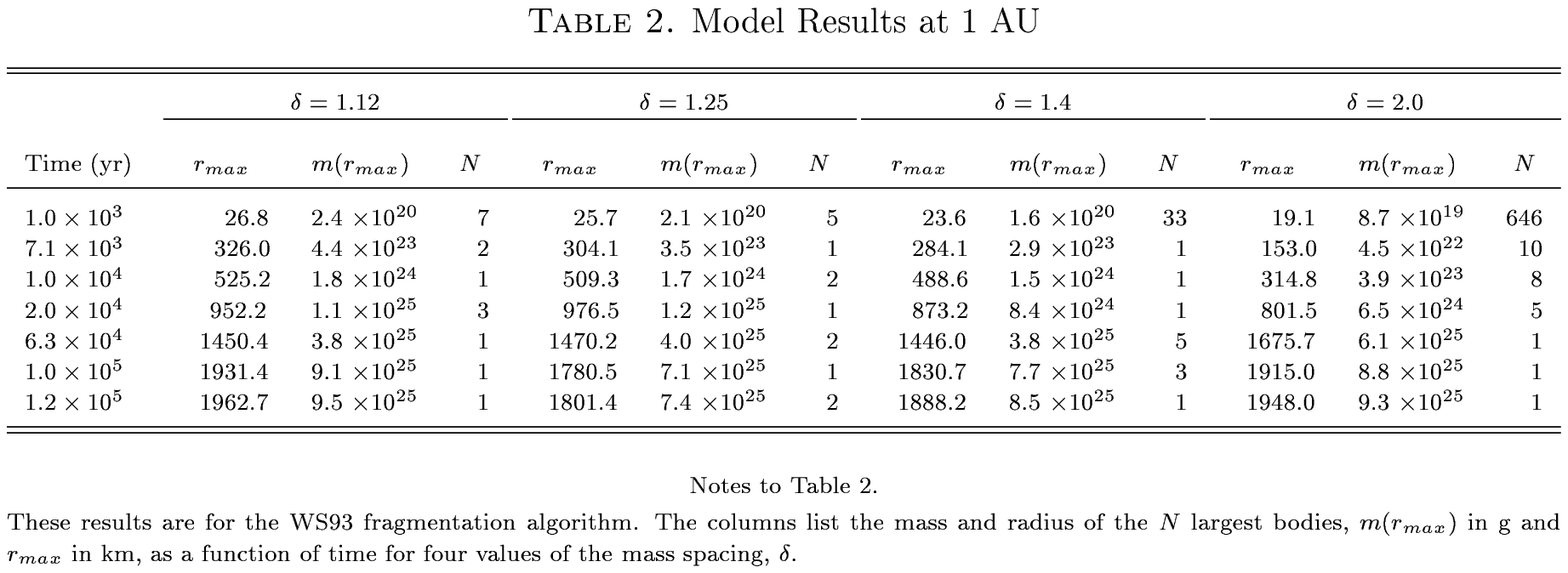}

\epsffile{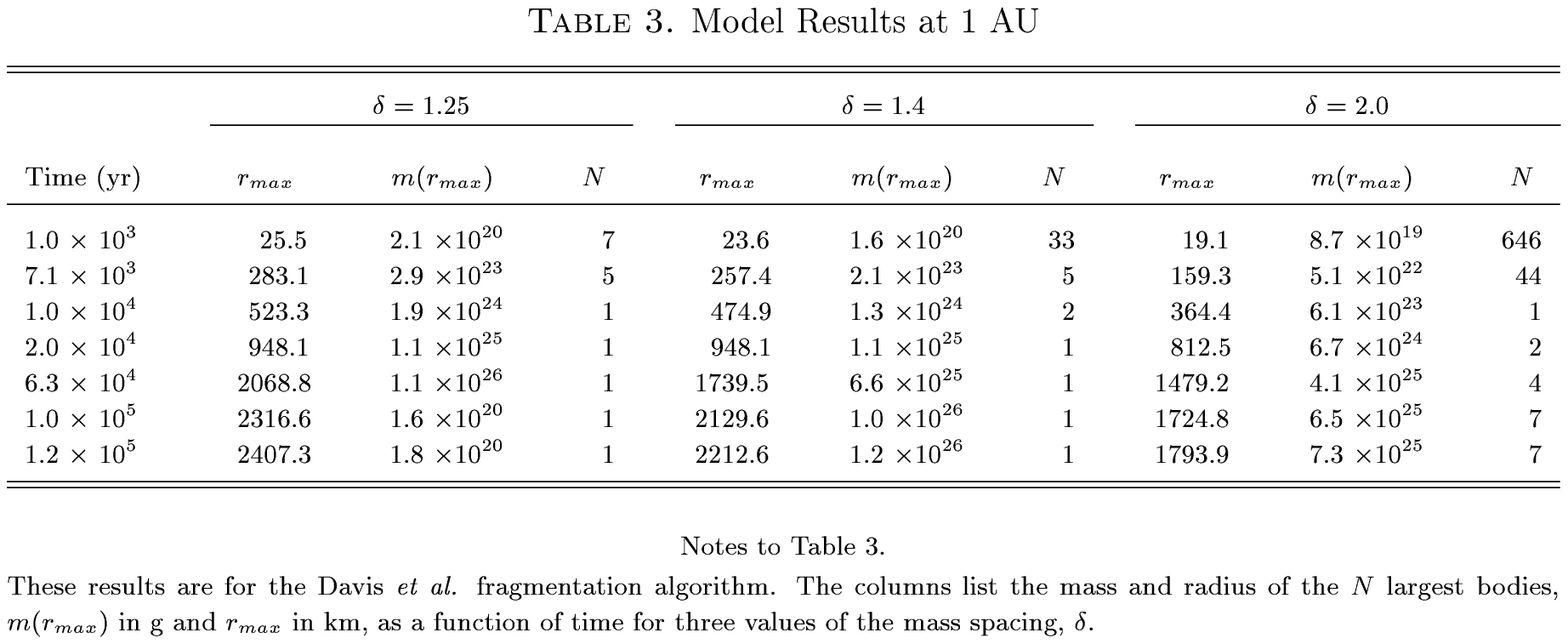}

\epsffile{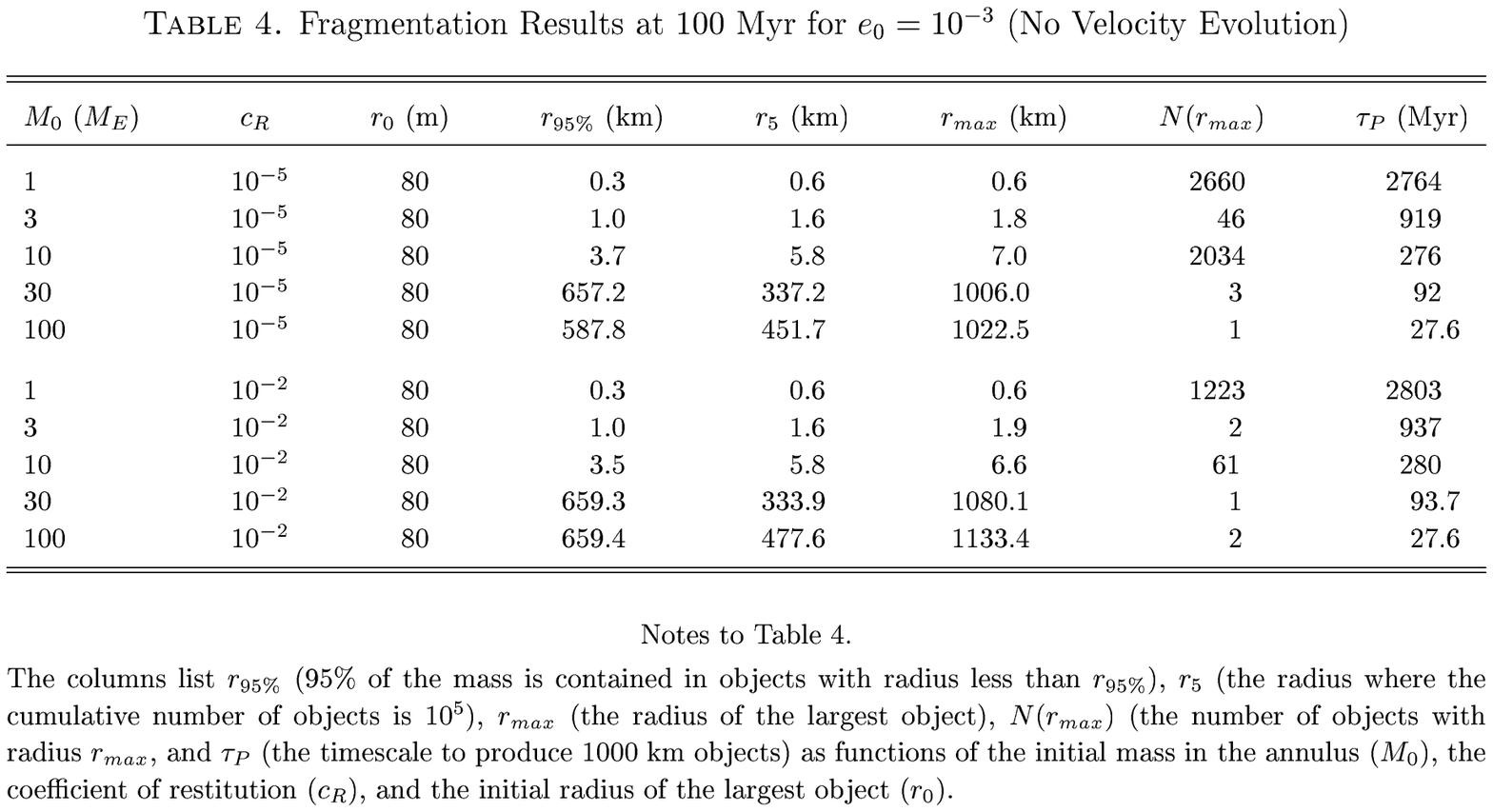}

\epsffile{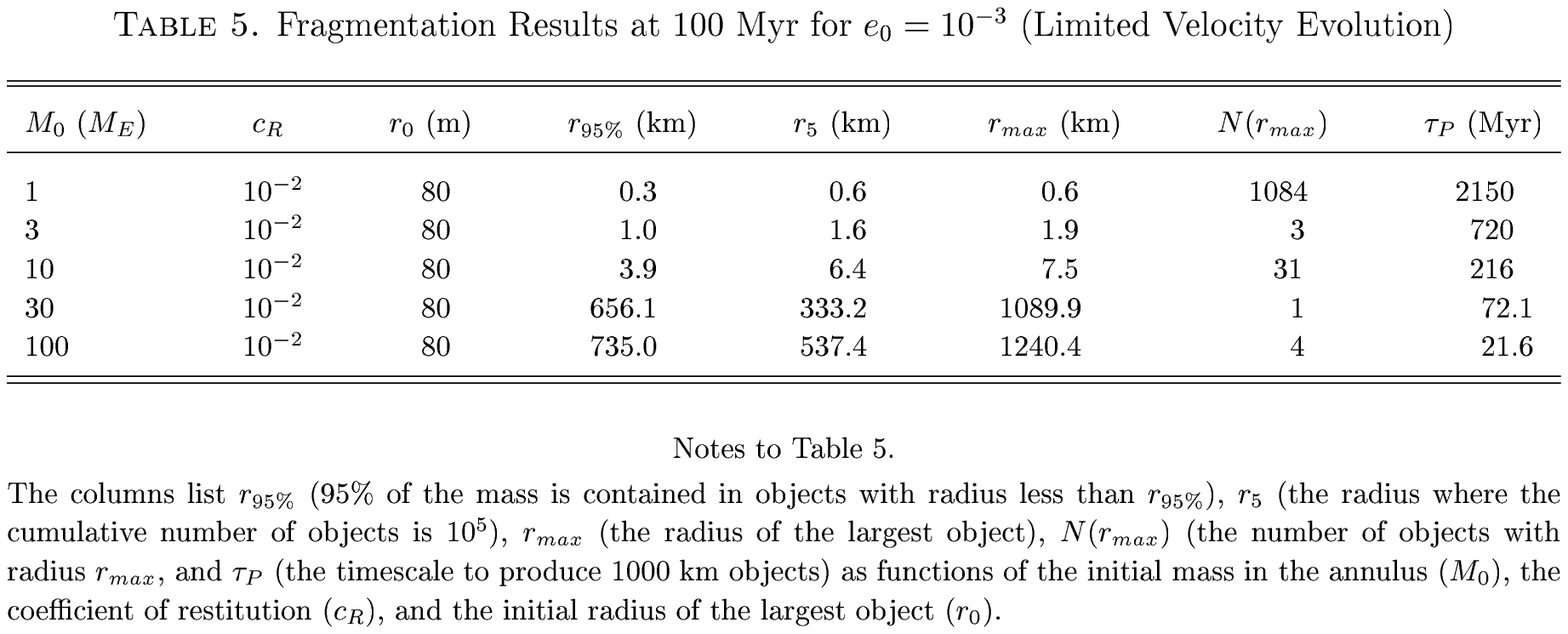}

\epsffile{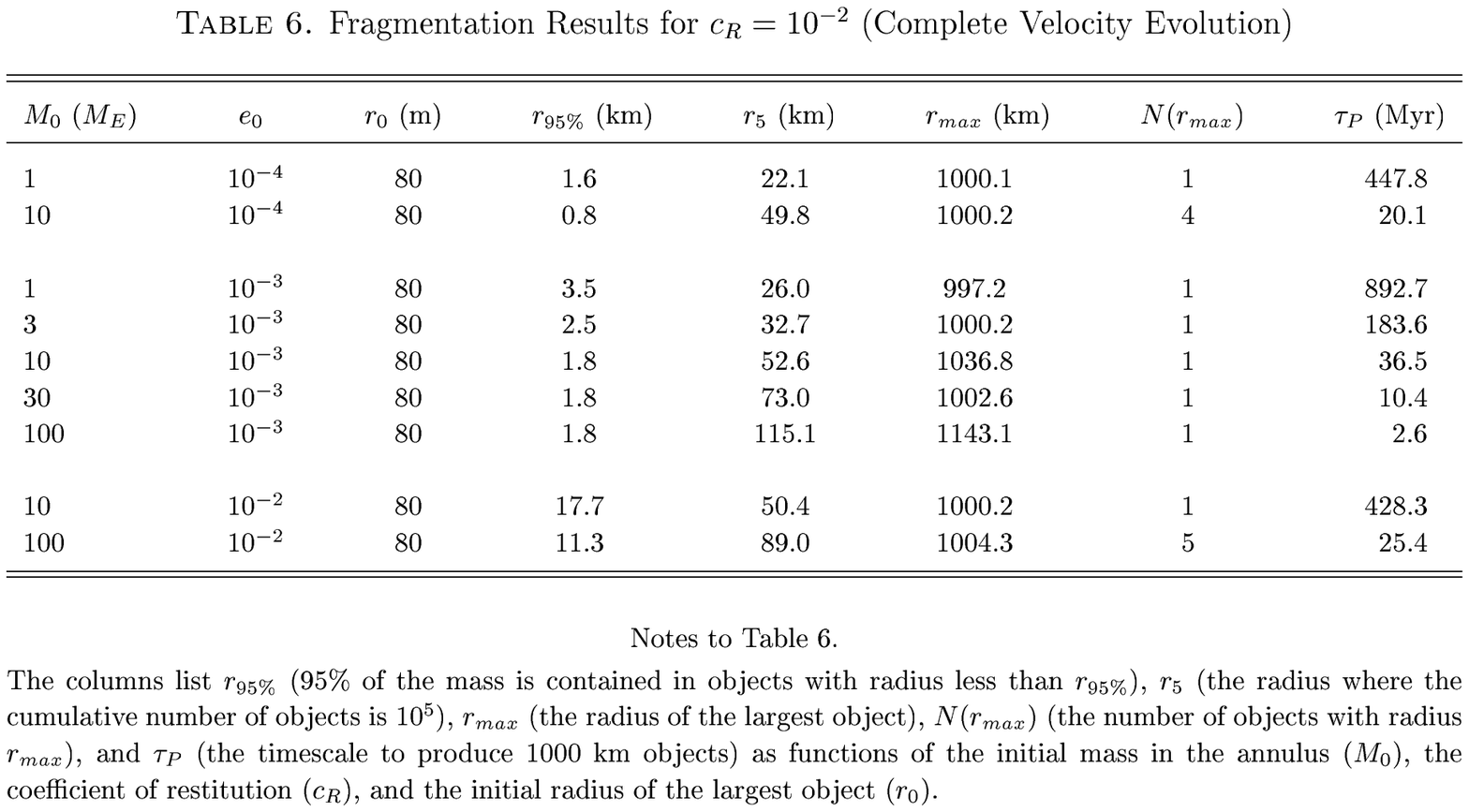}

\epsffile{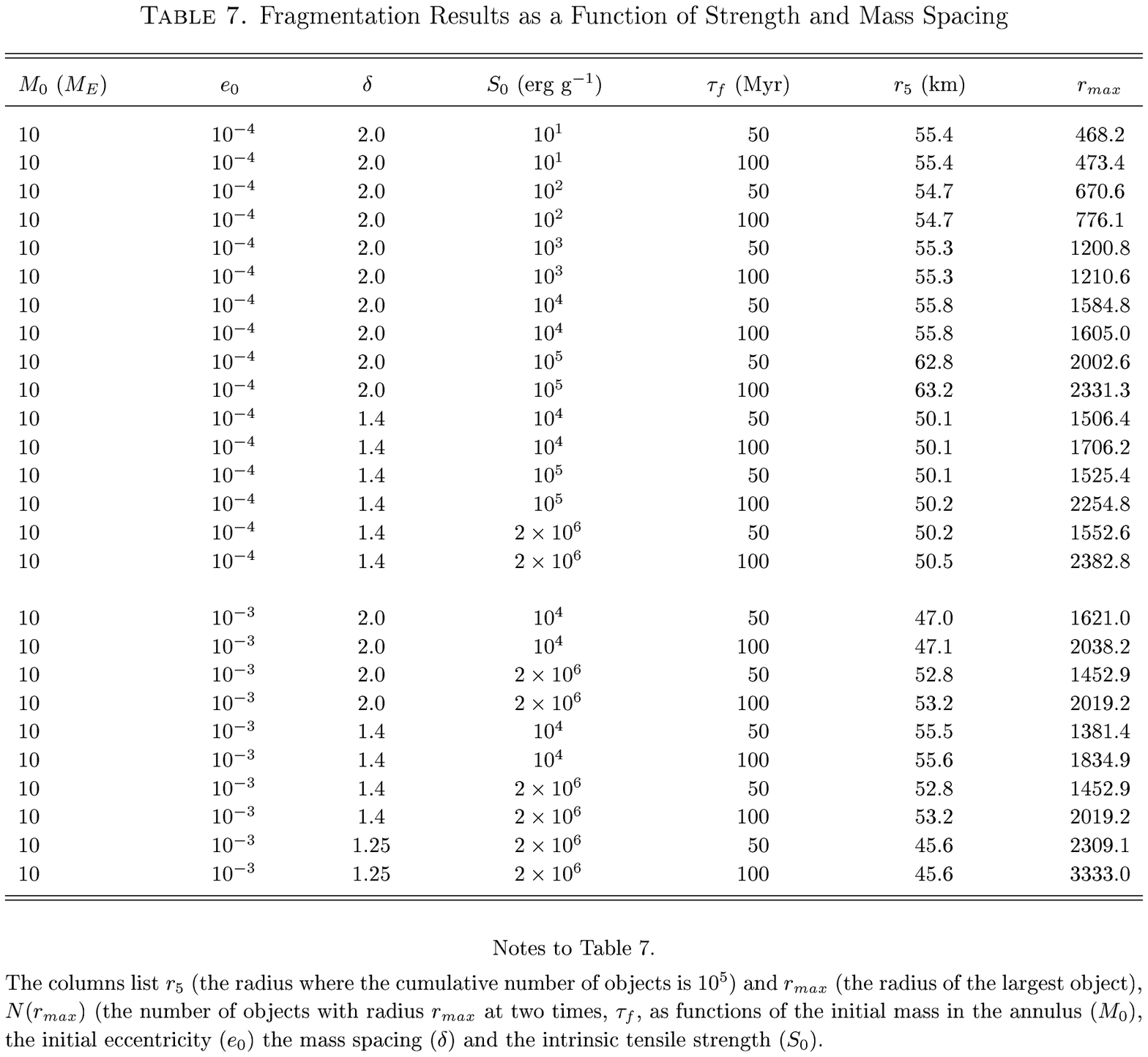}

\hskip -10ex
\epsfxsize=8.5in
\epsffile{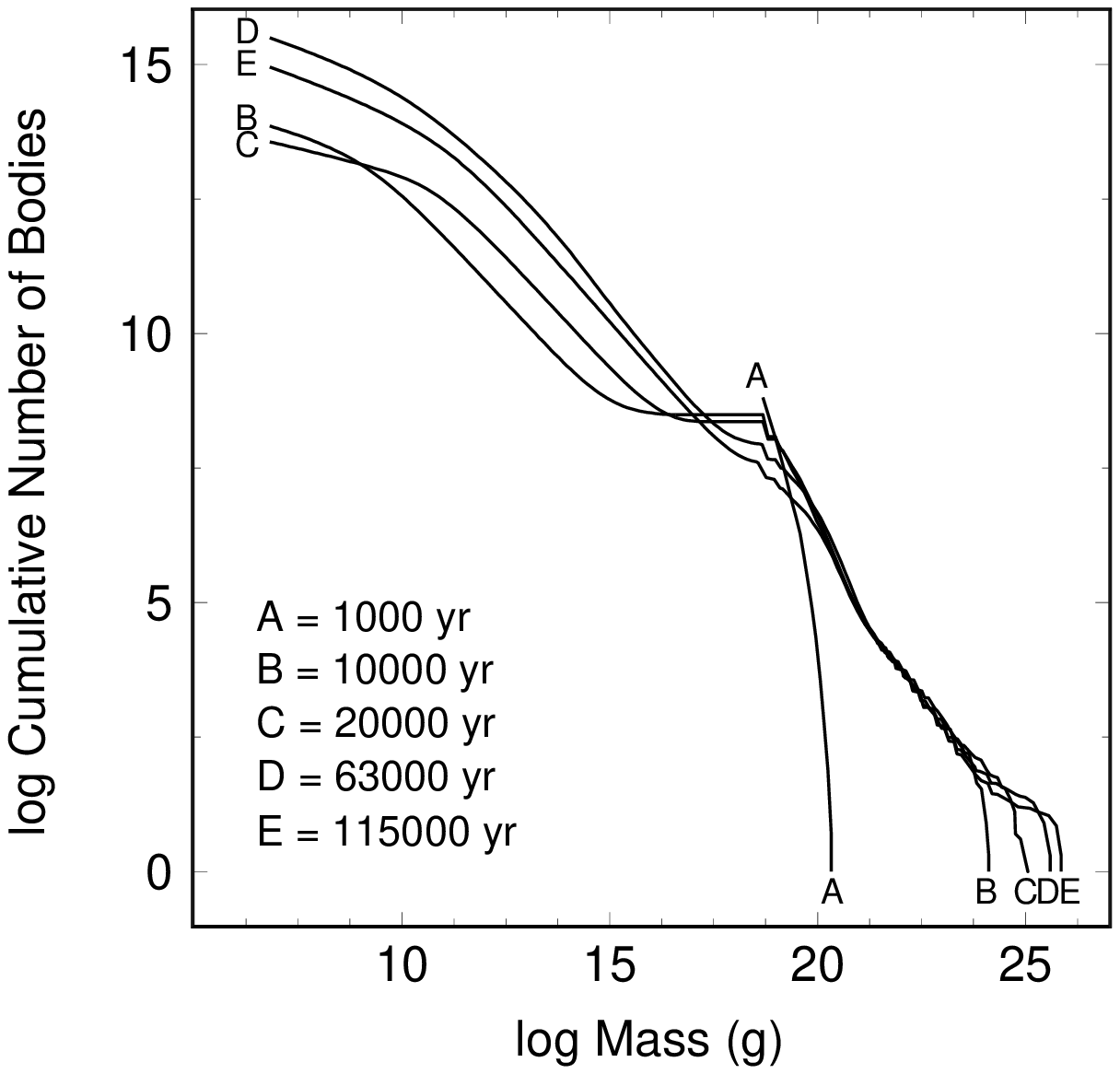}

\figcaption[Kenyon.fig1.ps]
{Cumulative mass distribution at selected times for a calculation
at 1 AU with $M_0$ = 0.667 $M_E$ and $\delta$ = 1.25.
A group of runaway bodies with $m_i \gtrsim 10^{24}$ g forms at
$\sim 10^4$ yr. These batches contain 13\% of the initial mass
at $6.3 \times 10^4$ yr and 17\% at $1.15~\times~10^5$ yr.}

\epsfxsize=7in
\hskip -5ex
\epsffile{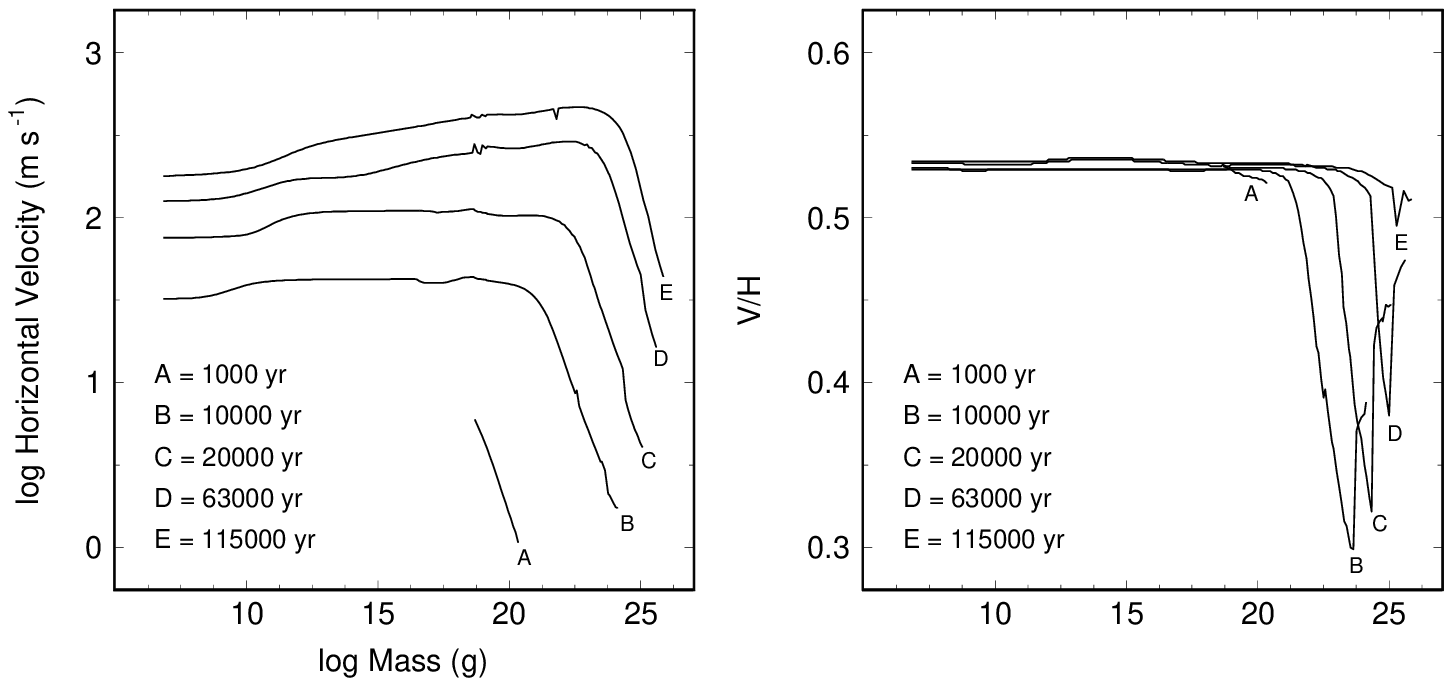}

\figcaption[Kenyon.fig2.ps]
{Velocity evolution for the 1 AU model in Fig. 1.
(a) Horizontal velocity distribution.
Viscous stirring increases all velocities with time;
dynamical friction brakes the runaway bodies and
increases velocities of the lowest mass bodies.
(b) Ratio of vertical to horizontal velocity.
The ratio remains close to the equilibrium value of
V/H = 0.53 for all but the most massive objects,
where V/H $\sim$ 0.3--0.5.}

\epsfxsize=8.0in
\epsffile{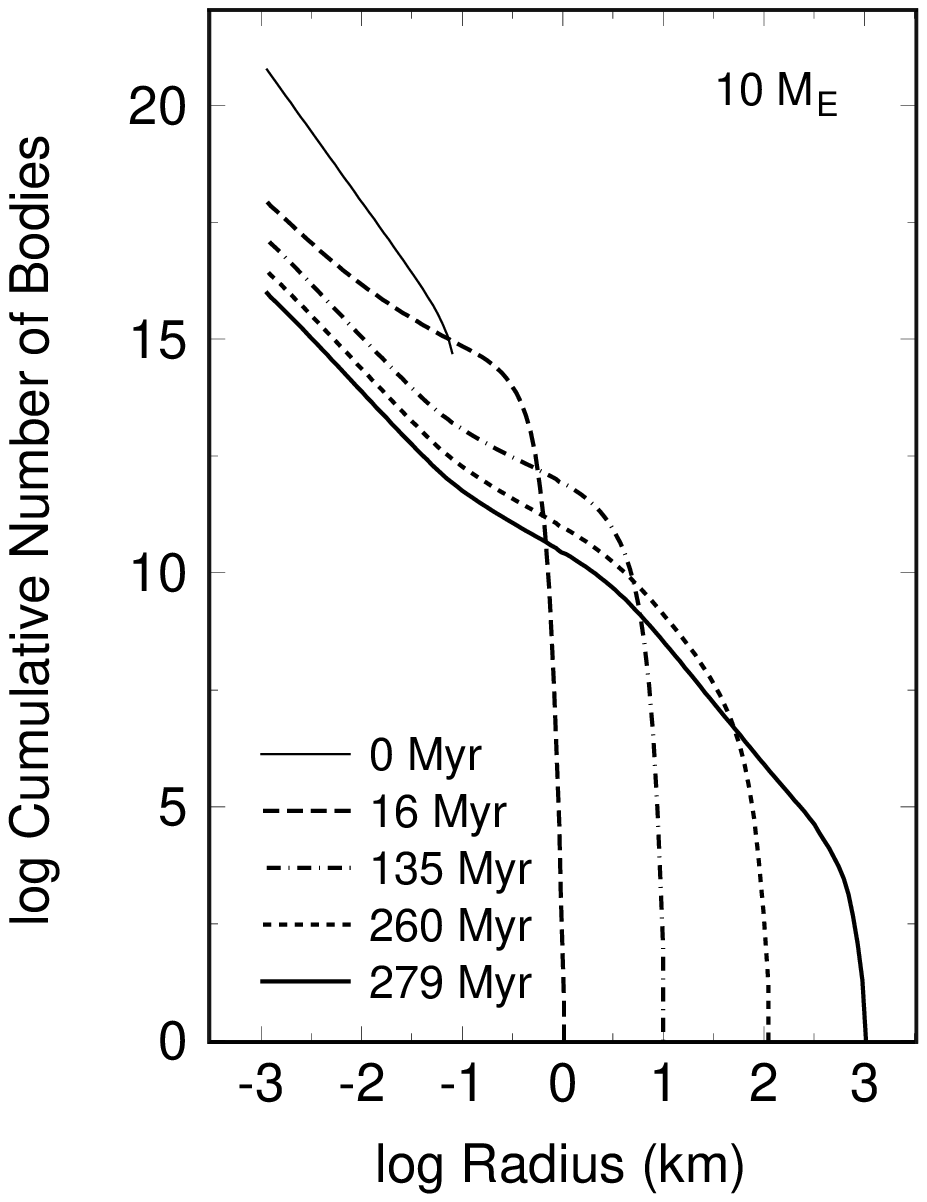}

\figcaption[Kenyon.fig3.ps]
{Cumulative size distributions for a Kuiper Belt model with
$M_0$ = 10 $M_E$ and $c_1$ = $c_2$ = $10^{-5}$.  The
eccentricity is constant in time at $e = 10^{-3}$.
Collisional growth is slow for 200 Myr until the largest
bodies have $r_{max}$ = 50 km.  Runaway growth begins when $r_{max}
\simgreat $ 100 km. These bodies then grow to sizes of $10^3$ km to
$10^4$ km in 20--30 Myr.  The size distribution consists of a
fragmentation tail with $N_c \propto r_i^{-2.25}$ for $r_i \lesssim$ 0.1 km,
a transition region with a flatter size distribution, and an
accretion component with $N_C \propto r_i^{-2.6}$.}

\epsfxsize=8.0in
\epsffile{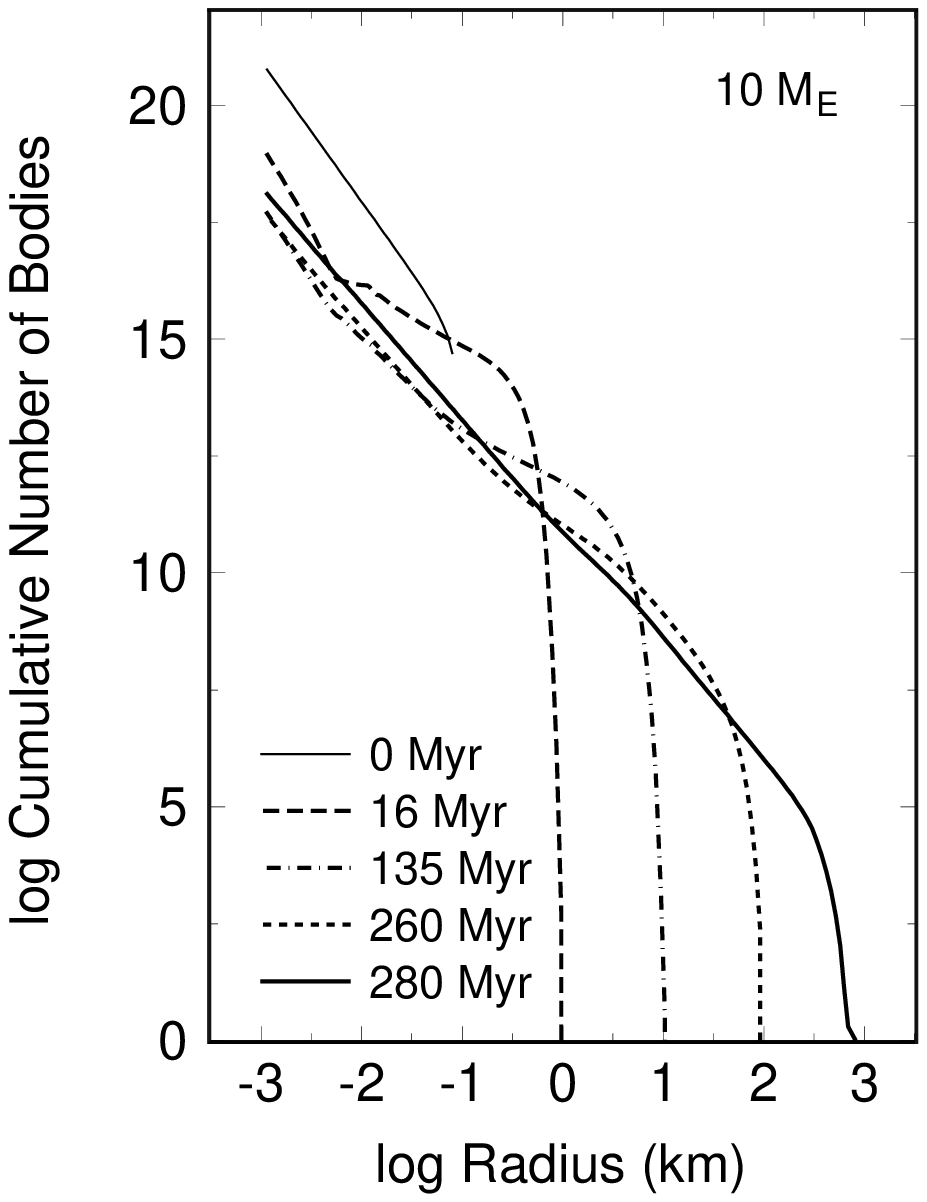}

\figcaption[Kenyon.fig4.ps]
{Cumulative size distributions for a Kuiper belt model with
$M_0$ = 10 $M_E$, $c_1$ = $10^{-2}$, and $c_2$ = $10^{-3}$.  The
eccentricity is constant in time at $e_0 = 10^{-3}$.
The evolution is identical to that in Figure 3, except for
the appearance of a `kink' in the size distribution due to rebounds at
$r_i \approx$ 5~m.  As the evolution proceeds, debris from collisions
with $r_i \gg$ 5~m suppresses this kink and the size distribution more
closely follows the $N_c \propto r_i^{-2.5}$ derived in calculations
without rebounds. }

\epsfxsize=8.0in
\epsffile{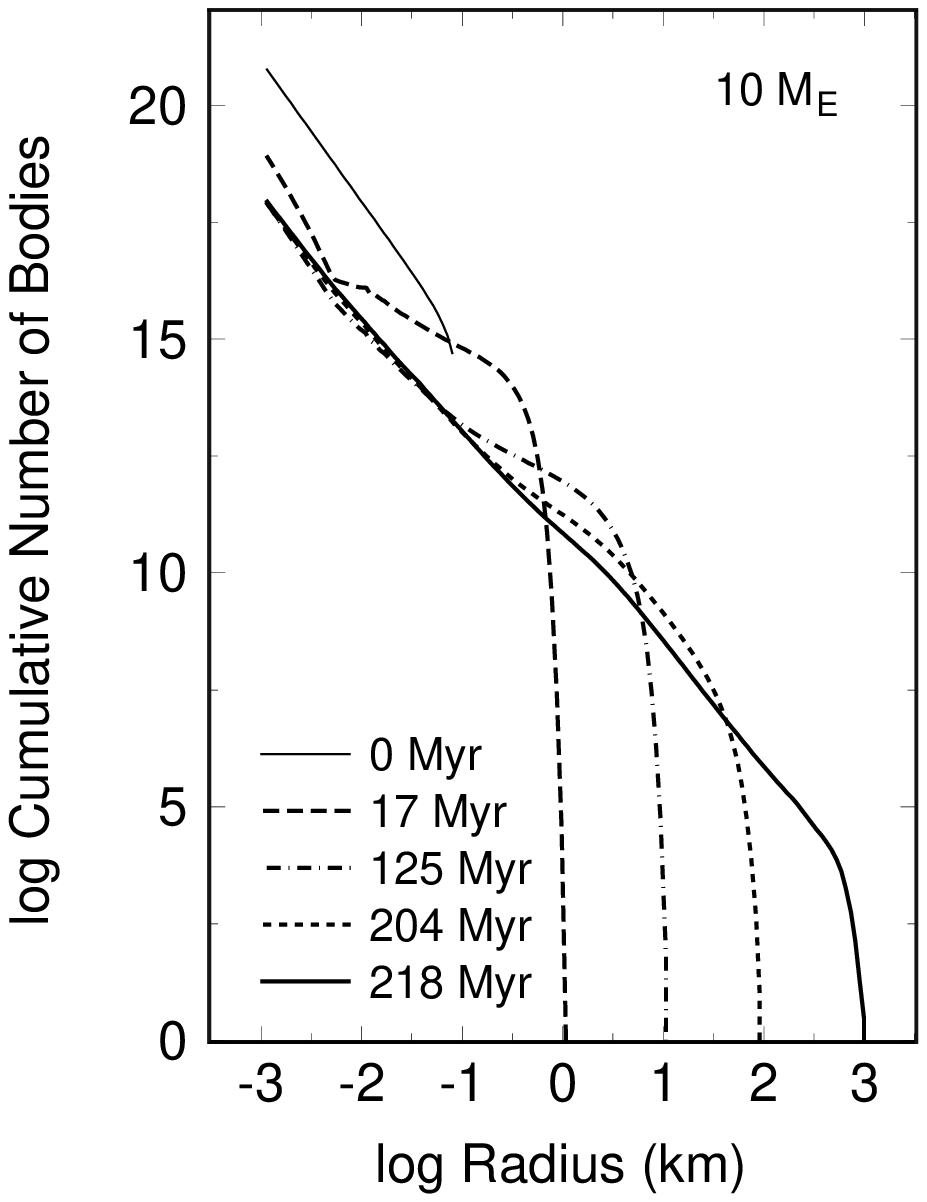}

\figcaption[Kenyon.fig5.ps]
{Cumulative size distributions for a Kuiper belt model with
$M_0$ = 10 $M_E$, $c_1$ = $10^{-2}$,
$c_2$ = $10^{-3}$, and limited velocity evolution.  Eccentricity changes
are due to fragmentation, which places a fixed fraction $f_{KE}$
of the kinetic energy of each collision into the debris.
The size distribution
develops a kink at $r_i \approx$ 5 m due to rebounds at 10--50 Myr
when $r_{max} \approx$ 1--3 km.  As the maximum radius increases,
the size distribution approaches a smooth power law distribution,
$N_C \propto r_i^{-2.5}$ from $r_i$ = 1 m up to $r_i$ = 300--500 km.}

\epsfxsize=8.0in
\epsffile{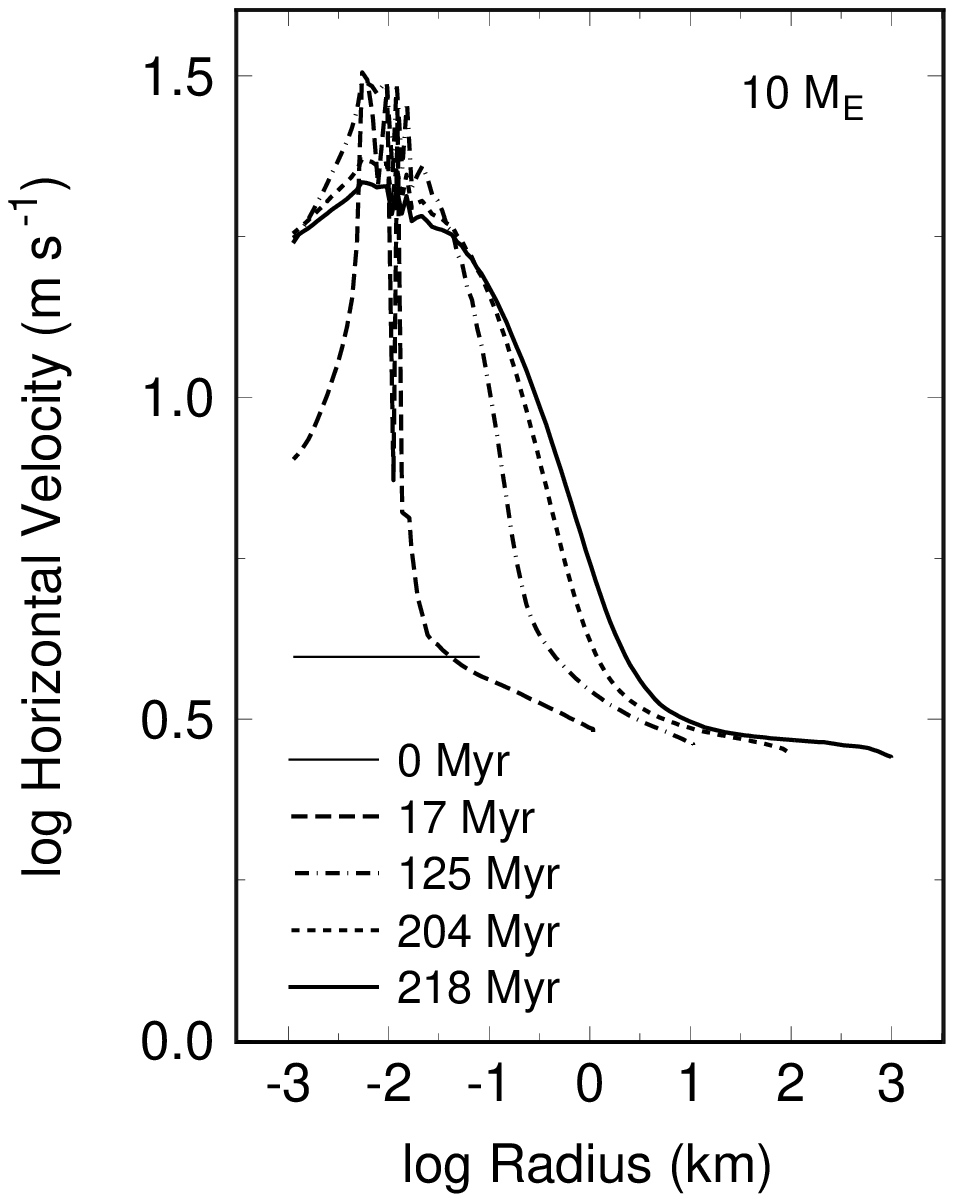}

\vskip -3ex
\figcaption[Kenyon.fig6.ps]
{Horizontal velocity distributions as a function of time for the limited
velocity evolution model in Figure 5.  The velocity dispersion of
all bodies begins at $h_i$ = 4 m s$^{-1}$.  As the maximum radius
increases from $r_{max}$ = 80 m to $r_{max}$ = 1 km, collision
debris increases the velocity of low mass bodies.  This increase
is most pronounced for bodies with $r_i \sim$ 10 m at early times,
because rebounds reduce the velocities of the lowest mass objects
(dotted line).  The high velocity dispersion propagates to bodies
with $r_i$ = 1--100 m as rebounds become less frequent and as
mergers of the largest bodies produce more debris (dot-dashed
and dashed lines). Once $r_{max} \approx$ 1000 km, the velocity
distribution has settled into a high velocity component at $r_i \sim$
1--100 m, a transition component at $r_i \sim$ 0.1--3 km, and a
group of large objects, $r_i \gtrsim$ 3--5 km with $h_i \approx$
3 m s$^{-1}$.}

\epsfxsize=8in
\hskip -10ex
\epsffile{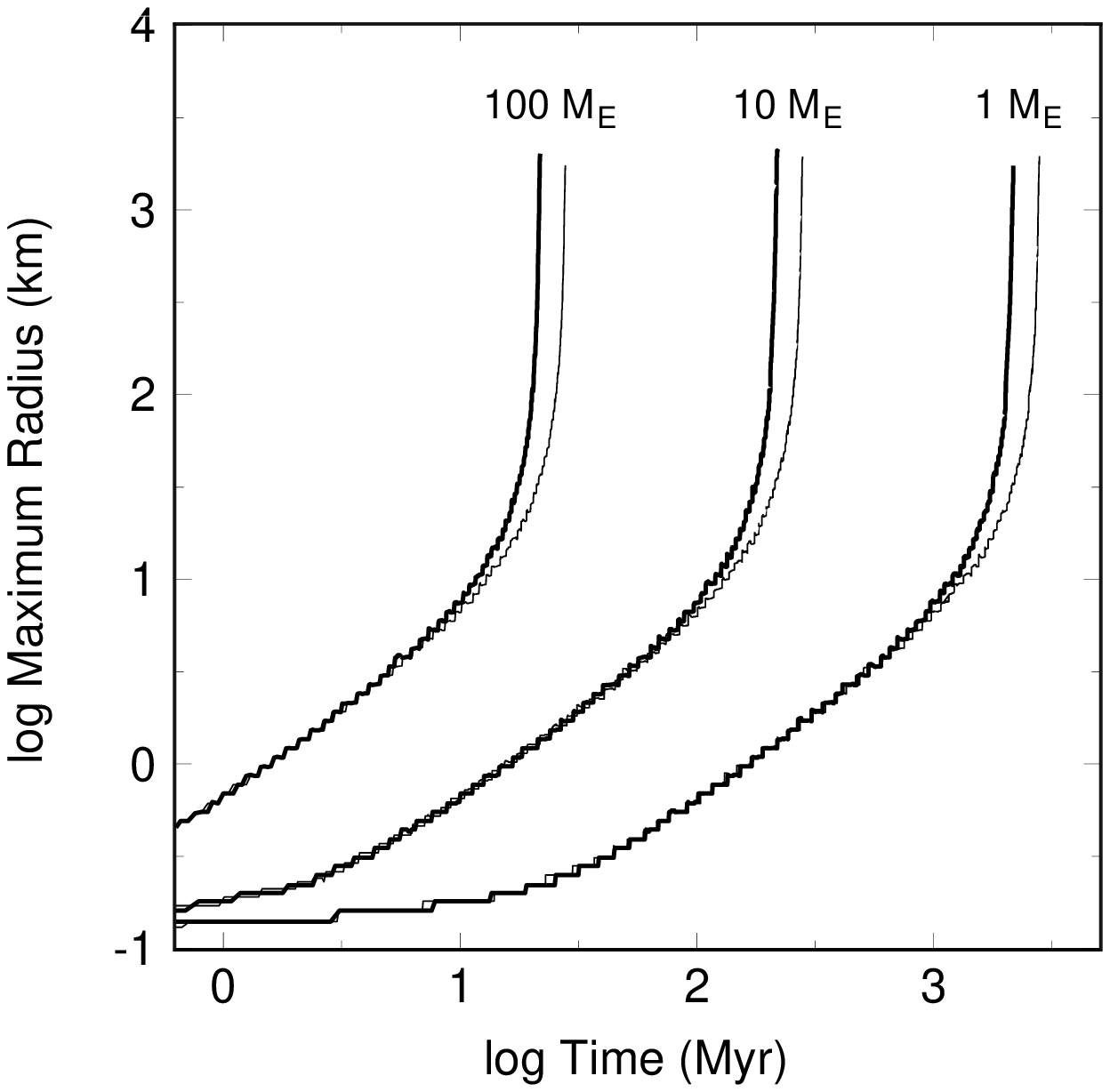}

\figcaption[Kenyon.fig7.ps]
{Maximum radius as a function of initial mass for models with
no velocity evolution (thin solid lines) and limited velocity
evolution (thick solid lines).  The initial eccentricity is $e_0 = 10^{-3}$.
The time to produce 1000 km objects scales inversely with initial mass,
$\tau_P \approx \tau_0 ~ (M_0 / 10 M_E)^{-1}$,
with $\tau_0 \approx $ 276--280 Myr for models with no velocity evolution
and $\tau_0 \approx $ 216 Myr for models with limited velocity evolution.
Fragmentation speeds up the growth of models with limited velocity
evolution by redistributing kinetic energy from large objects
to small objects.}

\epsfxsize=6.5in
\hskip -5ex
\epsffile{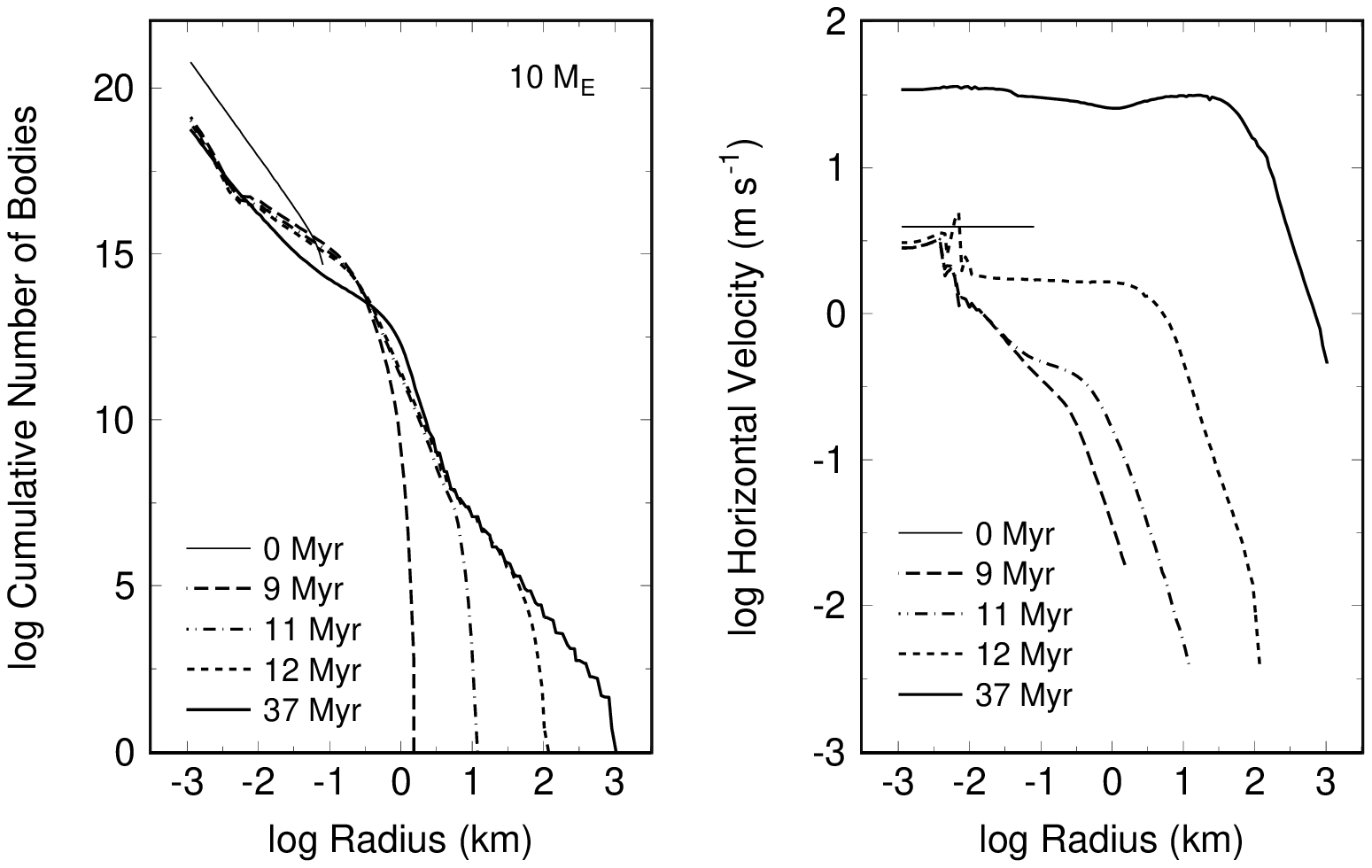}

\figcaption[Kenyon.fig8.ps]
{Size and velocity distributions for a model with $M_0$ = 10 $M_E$,
$e_0 = 10^{-3}$, $S_0 = 2 \times 10^6$ erg g$^{-1}$, and velocity
evolution: (a) cumulative size distribution, and (b)
horizontal velocity as a function of time. Collisional growth is
slow until the largest bodies have $r_{max}$ = 1--2 km
at 9--10 Myr. Collisional damping reduces the velocities of all
bodies to $\sim$ 1--2 m s$^{-1}$ on this timescale; dynamical
friction reduces the velocities of larger bodies to $\sim 10^{-2}$
m s$^{-1}$.  Runaway growth then produces objects with radii
of 100 km in another 2--3 Myr.  Viscous stirring increases
particle velocities as objects grow to sizes of 100--300 km,
and runaway growth ends.  A prolonged linear growth phase leads
to the production of 1000 km objects; the horizontal velocities
are then $\sim$ 30--40 m s$^{-1}$ for the smallest objects and
$\sim$ 1 m s$^{-1}$ for the largest objects.}

\epsfxsize=8in
\hskip -7ex
\epsffile{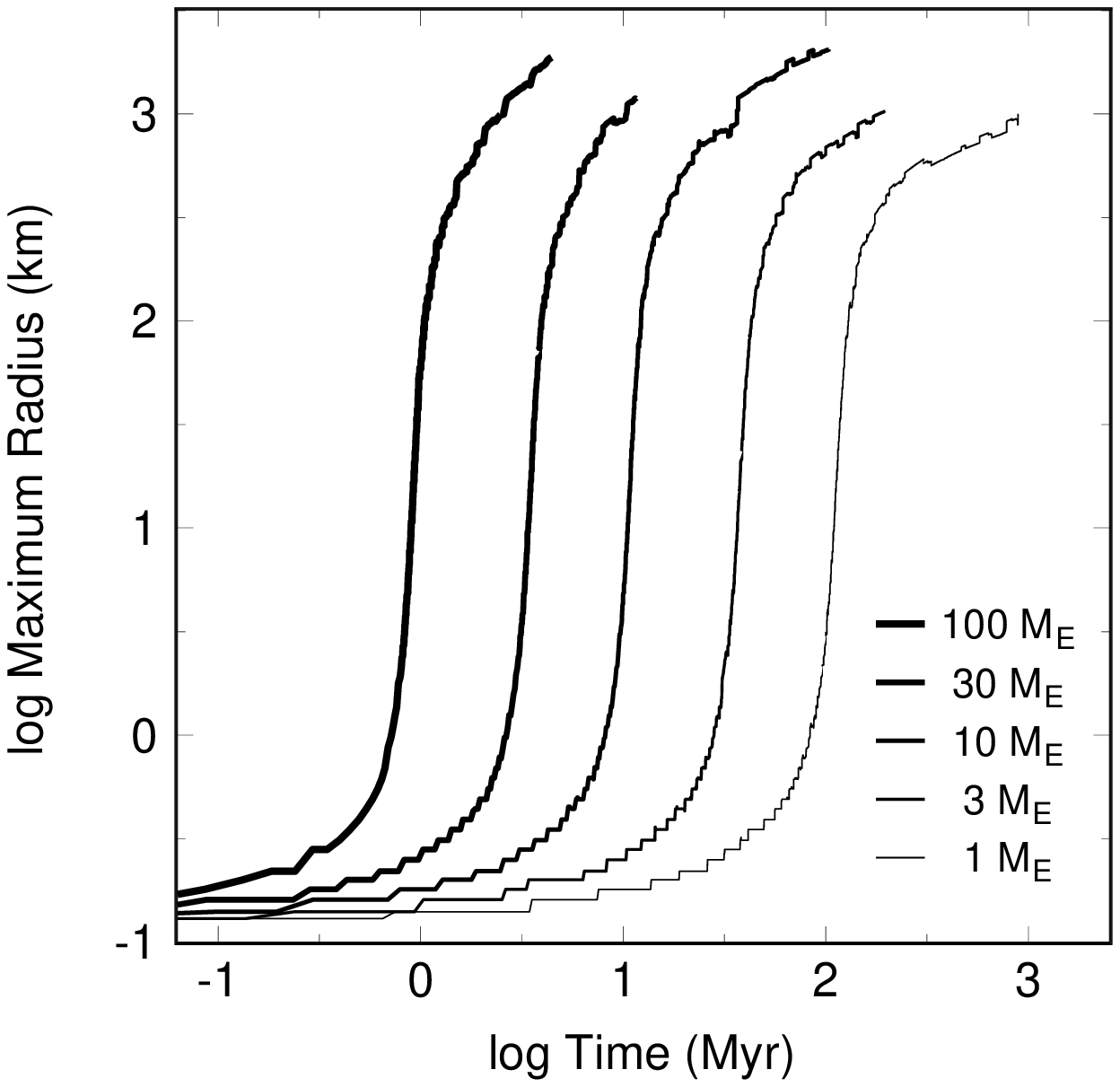}

\figcaption[Kenyon.fig9.ps]
{Evolution of the maximum radius, $r_{max}$, with time as a function of
initial mass, $M_0$, for $e_0 = 10^{-3}$.  The timescale
to reach runaway growth decreases with increasing $M_0$.}

\epsfxsize=8in
\hskip -7ex
\epsffile{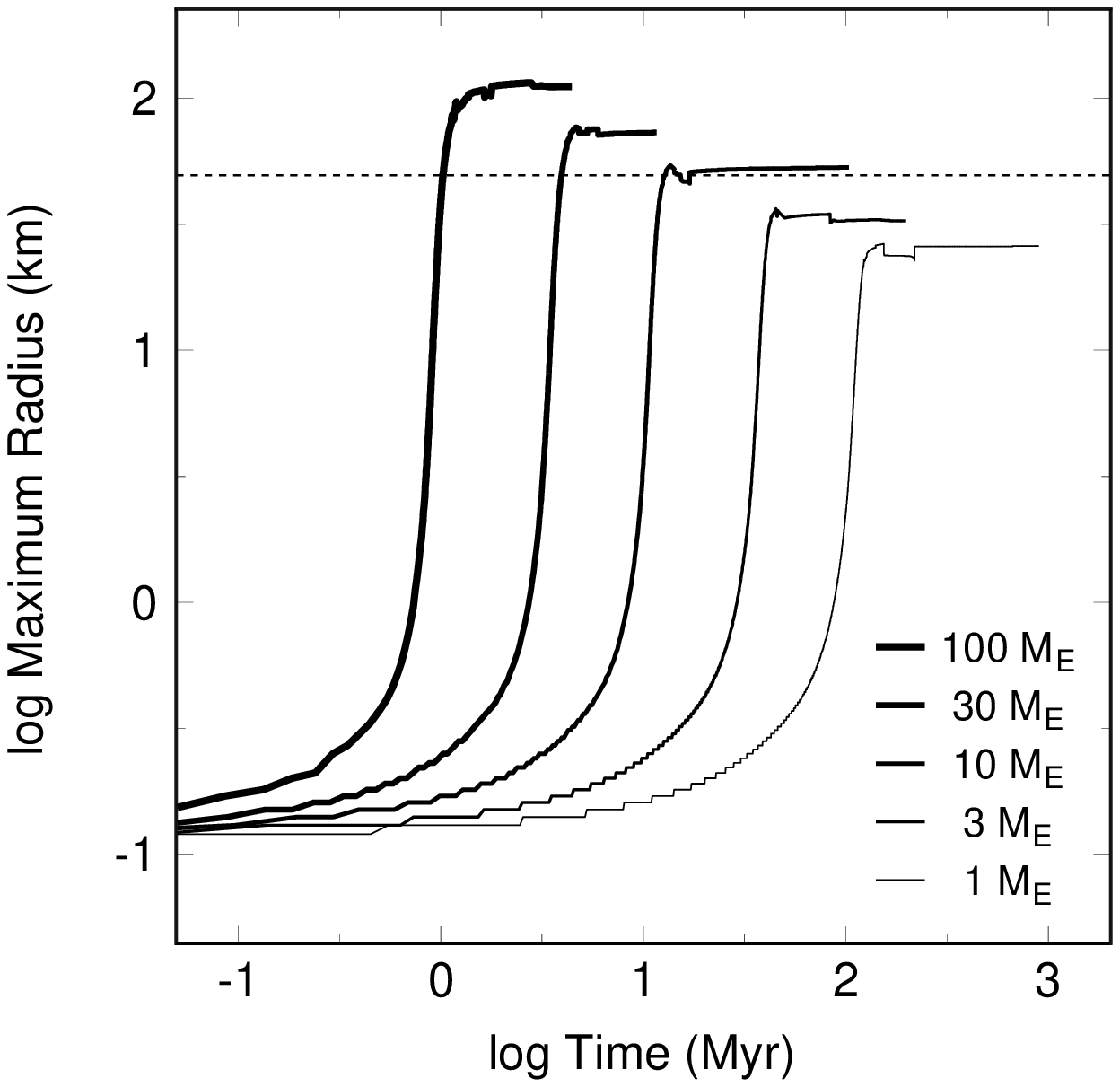}

\figcaption[Kenyon.fig10.ps]
{Evolution of $r_5$, the radius where the cumulative number of
objects is $10^5$, with time as a function of initial mass,
$M_0$, for $e_0 = 10^{-3}$.  The constraint on $r_5$ set
by current observations is indicated by the horizontal dashed line.}

\epsfxsize=9in
\epsffile{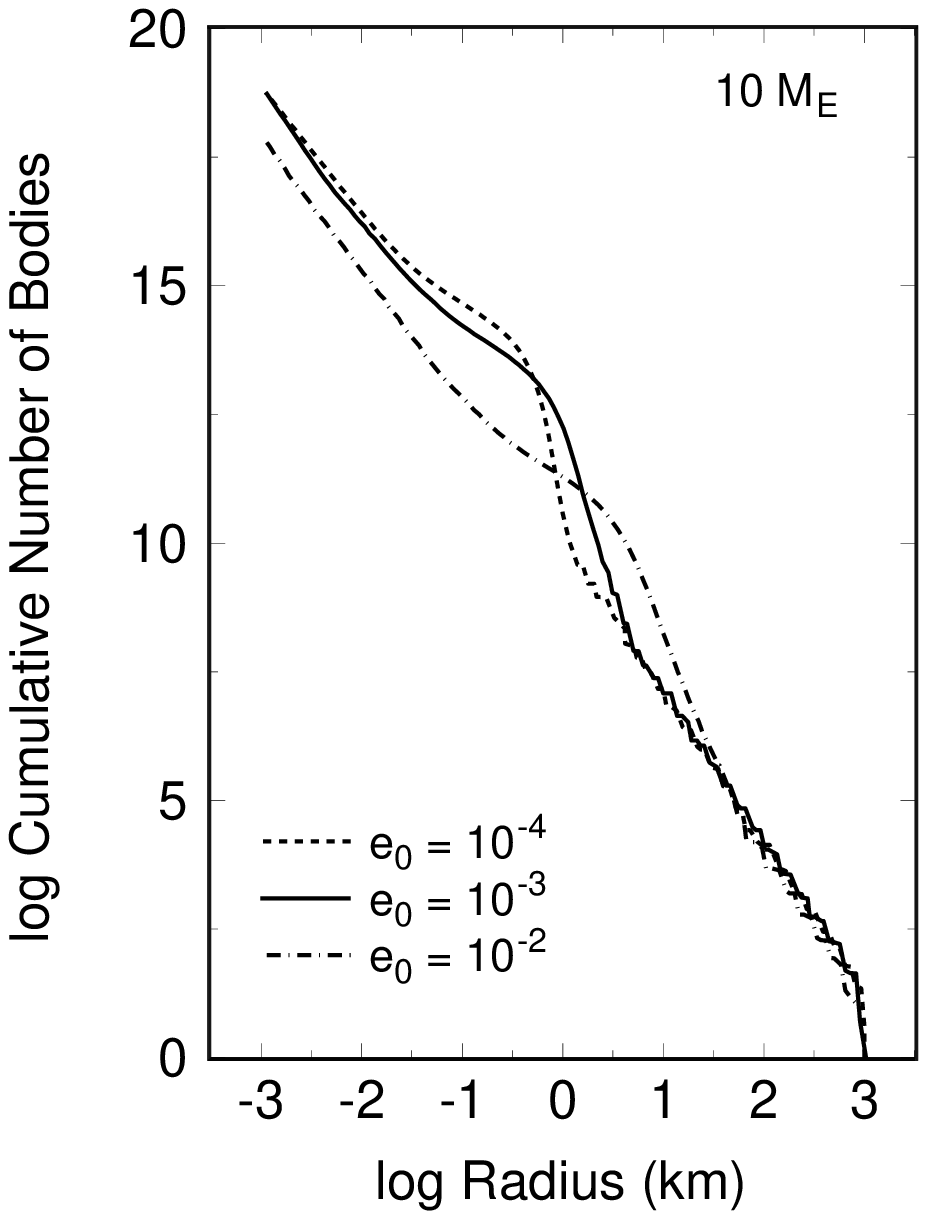}

\figcaption[Kenyon.fig11.ps]
{Cumulative size distribution when $r_{max}$ = 1000 km for three values
of the initial eccentricity, $e_0$.  Each size distribution follows
two power laws, $N_C \propto r^{-2.5}$ at small radii and
$N_C \propto r^{-3}$ at large radii.  The transition between
the two power laws moves to larger radii as $e_0$ increases.}

\epsfxsize=9in
\epsffile{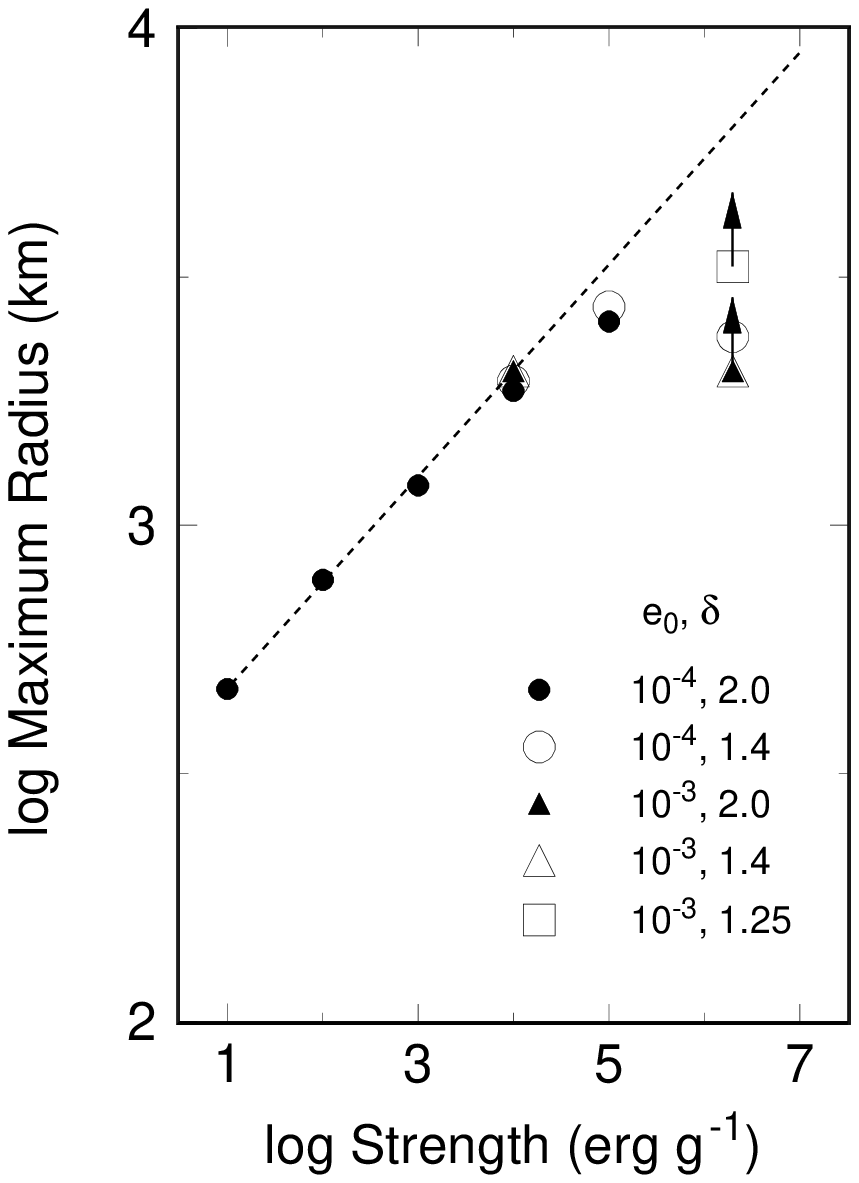}

\figcaption[Kenyon.fig12.ps]
{Maximum radius as a function of strength for models with
$M_0 = 10~M_E$ and various $e_0$ and $\delta$ as listed
in the legend.}

\epsfxsize=9in
\epsffile{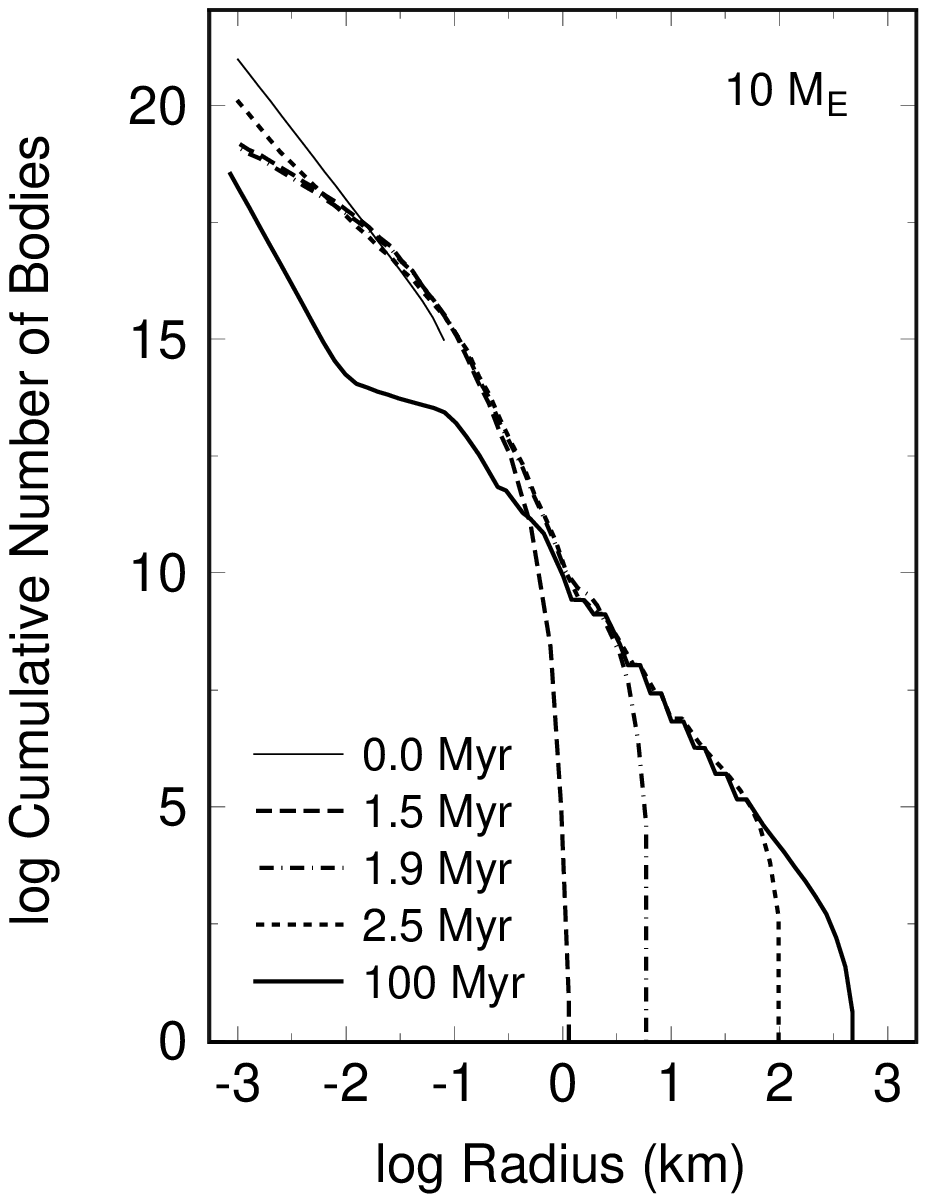}

\figcaption[Kenyon.fig13.ps]
{Cumulative size distributions for a model with
$M_0 = 10~M_E$, $e_0 = 10^{-4}$, and $S_0$ = 100 erg g$^{-1}$.
Runaway growth produces objects with $r_{max} \approx$ 100 km as in
the standard model with $S_0$ = $3 \times 10^6$ erg g$^{-1}$.
Catastrophic disruption begins when $r_{max} \approx$ 400 km.
Continued mass loss due to catastrophic disruption reduces the
population of low mass objects with $r_i \lesssim$ 0.1 km, which
are completely removed from the distribution for $\tau \gtrsim$ 100 Myr.}


\begin{thebibliography}{}

\bibitem[Akeson \etal 1998]{ake98} Akeson, R. L., Koerner, D. W.,
\& Jensen, E. L. N. 1998, ApJ, 505, 358

\bibitem[Bailey 1994]{bai94} Bailey, M. 1994,  
In Asteroids, Comets, Meteors 1993,
eds. A. Milani, M. DiMartino, and A. Cellino (Kluwer Academic
Publishers, Dordrecht), pp. 443 - 459.

\bibitem [Backman \& Paresce 1993]{bac93} 
Backman, D. E., \& Paresce, F. 1993, in
{\it Protostars and Planets III}, eds. E. H. Levy \& J. I. Lunine,
Tucson, Univ of Arizona, p. 1253

\bibitem [Backman \etal 1995]{bac95} Backman, D. E., Dasgupta, A., \& 
Stencel, R. E. 1995, ApJ, 450, L35

\bibitem[Barge \& Pellat 1990]{bar90} Barge, P., \& Pellat, R. 1990,
Icarus, 85, 481

\bibitem[Barge \& Pellat 1991]{bar91} Barge, P., \& Pellat, R. 1991,
Icarus, 93, 270

\bibitem[Barge \& Pellat 1993]{bar93} Barge, P., \& Pellat, R. 1993,
Icarus, 104, 79

\bibitem[Beckwith \& Sargent 1996]{bec96} Beckwith, S. V. W., \&
Sargent, A. I. 1996, Nature, 383, 139

\bibitem[Beckwith \etal 1990]{bec90} Beckwith, S. V. W., Sargent, A. I., 
Chini, R., \& G\"usten, R.  1990, AJ, 99, 924

\bibitem[Boss 1997]{bos97} Boss, A. P. 1997, Science, 276, 1836

\bibitem[Close \etal 1997]{clo97} Close, L. M., Roddier, F., Hora, J. L.,
Graves, J. E., Northcott, M., Roddier, C., Hoffman, W. F., Dayal, A.,
Fazio, G. G., \& Deutsch, L. K. 1997, ApJ, 489, 210

\bibitem[Cuzzi \etal 1993a]{cz93a} Cuzzi, J. N., Dobrovolskis, A. R.,
\& Champney, J. M. 1993, Icarus, 100, 102

\bibitem[Cuzzi \etal 1993b]{cz93b} Cuzzi, J. N., Dobrovolskis, A. R.,
\& Champney, J. M. 1993, Icarus, 106, 102

\bibitem[Davis \etal 1985]{dav85} Davis, D. R., Chapman, C. R., 
Weidenschilling, S. J., \& Greenberg, R. 1985, Icarus, 62, 30

\bibitem[Davis \& Farinella 1997]{dav97} Davis, D. R., \& 
Farinella, P. 1997.  Icarus, 125, 50

\bibitem[Davis \etal 1994]{dav94} Davis, D. R., Ryan, E. V., \& 
Farinella, P. 1994, Planet. Space Sci., 42, 599

\bibitem[Davis \etal 1989]{dav89} Davis, D. R., Weidenschilling, S. J.,
Farinella, P., Paolicchi, P., \& Binzel, R. P. 1989, in
{\it Asteroids II,} edited by R. P. Binzel, T. Gehrels, \&
M. S. Matthews, Tucson, Univ. of Arizona Press, p. 805

\bibitem[Dohnanyi 1969]{doh69} Dohnanyi, J. W. 1969, J. Geophys. Res., 74, 2531

\bibitem[Duncan \etal 1995]{dun95} Duncan, M. J., Levison, H. F., 
\& Budd, S. M. 1995,  AJ, 110, 3073

\bibitem[Durda \etal 1998]{dur98} Durda, D. D., Greenberg, R., \& Jedicke, P.
1998, Icarus, 135, 431

\bibitem[Farinella \etal 1982]{far82} Farinella, P., Paolicchi, P., \&
Zappal\'a, V. 1992, A\&A, 253, 604

\bibitem[Fern\'andez \& Ip 1981]{fer81} Fern\'andez, J. A., \& 
Ip, W.-H. 1981, Icarus, 47, 470

\bibitem[Fern\'andez \& Ip 1984]{fer84} Fern\'andez, J. A., \& 
Ip, W.-H. 1984, Icarus, 58, 109

\bibitem[Fujiwara 1980]{fuj80} Fujiwara, A. 1980, Icarus, 41, 356

\bibitem[Gault \etal 1963]{gau63} Gault, D. E., Shoemaker, E. M.,
\& Moore, H. J. 1963, NASA TN, 1767

\bibitem[Gladman \& Kavelaars 1997]{gla97} Gladman, B., \& Kavelaars, J. J. 
1997, A\&A, 317, L35

\bibitem[Gladman \etal 1998]{gla98} Gladman, B., Kavelaars, J. J.,
Nicholson, P. D., Loredo, T. J., \& Burns, J. A.  1998, AJ, 116, 2042

\bibitem[Goldreich \& Ward 1973]{gol73} Goldreich, P., \& 
Ward, W. R. 1973,  ApJ, 183, 1051

\bibitem[Greaves \etal 1998]{gre98} Greaves, J. S. \etal 1998, ApJ, 506, L133

\bibitem[Greenberg \etal 1978]{gre78}
Greenberg, R., Wacker, J. F., Hartmann, W. K., \& Chapman,
C. R. 1978, Icarus, 35, 1

\bibitem[Greenberg \etal 1984]{gre84} Greenberg, R., Weidenschilling, S. J.,
Chapman, C. R., \& Davis, D. R. 1984, Icarus, 59, 87

\bibitem[Hartmann \etal 1998]{har98} Hartmann, L., Calvet, N., Gullbring, E.,
\& D'Alessio, P. 1998, ApJ, 495, 385

\bibitem[Hayashi 1981]{hay81} Hayashi, C. 1981, Prog Theor Phys Suppl, 70, 35

\bibitem[Hogerheijde \etal 1997]{hoh97} Hogerheijde, M. R., 
van Langevelde, H. J., Mundy, L. G., Blake, G. A., \& van Dishoeck, E. F. 
1997, ApJ, 490, L99

\bibitem[Holman \& Wisdom 1993]{hol93}
Holman, M. J., \& Wisdom, J. 1993,  AJ, 105, 1987

\bibitem[Holsapple 1993]{hls93} Holsapple, K. A. 1993,
Ann. Rev. Earth Planet. Sci., 21, 333

\bibitem[Holsapple 1994]{hls94} Holsapple, K. A. 1994,
Planet. Space Sci., 42, 1067

\bibitem[Housen \& Holsapple 1990]{hou90} Housen, K., \& Holsapple, K.
1990, Icarus, 84, 226

\bibitem[Housen \etal 1991]{hou91} Housen, K., Schmidt, R. M., \& 
Holsapple, K.  1991, Icarus, 94, 180

\bibitem[Hornung \etal 1985]{hor85} Hornung, P., Pellat, R., \& Barge, P.
1985, Icarus, 64, 295

\bibitem[Ida 1990]{ida90} Ida, S. 1990, Icarus, 88, 129

\bibitem[Ip 1989]{ip89} Ip, W.-H. 1989, Icarus, 80, 167

\bibitem[Irwin \etal 1995]{ir95} Irwin, M., Tremaine, S., \& 
${\rm \dot{Z}ytkow}$, A. N. 1995, AJ., 110, 3082

\bibitem[Jayawardhana \etal 1998]{jay98} Jayawardhana, R. \etal 1998,
ApJ, 503, L79

\bibitem[Jedicke \& Metcalfe 1998]{jed98} Jedicke, P., \& Metcalfe, T. S.
1998, Icarus, 131, 245

\bibitem[Jewitt \& Luu 1993]{jew93} Jewitt, D., \& Luu, J. 1993, 
Nature, 362, 730

\bibitem[Jewitt \& Luu 1995]{jew95} Jewitt, D., \& Luu, J. 1995, AJ, 109, 1867

\bibitem[Jewitt \etal 1996]{jew96} Jewitt, D., Luu, J., \& Chen, J. 1996,
AJ, 112, 1225

\bibitem[Jewitt \etal 1998]{jew98} Jewitt, D., Luu, J. X., \& Trujillo, C.
1998, AJ, 115, 2125

\bibitem[KL98]{kl98} Kenyon, S. J., \& Luu, J. X.
1998, AJ, 115, 2136 (KL98) 

\bibitem[Kenyon \& Luu 1999]{kl99} Kenyon, S. J., \& Luu, J. X.
1999, ApJL, submitted

\bibitem[Kim \etal 1994]{kmh94} Kim, S.-H., Martin, P. G., \& Hendry, P. D.
1994, ApJ, 422, 164

\bibitem[Koerner \etal 1998]{koe98} Koerner, D. W., Ressler, M. E., 
Werner, M. W., \& Backman, D. E. 1998, ApJ, 503, L83

\bibitem[Kolvoord \& Greenberg 1992]{kol92} Kolvoord, R. A. 
\& Greenberg, R. 1992, Icarus, 98, 2

\bibitem[Lay \etal 1997]{lay97} Lay, O. P., Carlstrom, J. E., \& Hills, R. E.
1997, ApJ, 489, L917

\bibitem[Lazzaro \etal 1994]{laz94} Lazzaro, D., Sicardy, B., Roques, F.,
\& Greenberg, R. 1994, Icarus, 108, 59

\bibitem[Levison \& Duncan 1993]{lev93} Levison, H. F., \& 
Duncan, M. J. 1993,  ApJ, 406, L35

\bibitem[Li \& Greenberg 1997]{li97} Li, A., \& Greenberg, J. M. 1997, 
A\&A, 323, 566

\bibitem[Lissauer \& Stewart 1993]{lis93} Lissauer, J. J., \&
Stewart, G. R. 1993,  In {\it Protostars and Planets III,}
edited by E. H. Levy and J. I. Lunine, U. of Arizona Press, Tucson, 1061

\bibitem[Lissauer \etal 1996]{lis96}
Lissauer, J. J., Pollack, J. B., Wetherill, G. W., \& Stevenson,
D. J. 1996.  "Formation of the Neptune System."  In Neptune and
Triton, eds. D. P. Cruikshank, M. S. Matthews, and A. M. Schumann
(U. of Arizona Press, Tucson, pp. 37 - 108.

\bibitem[Luu \& Jewitt 1998]{luu98} Luu, J. X., \& Jewitt, D. 
1998, ApJ, 502, L91

\bibitem[Luu \etal 1997]{luu97} Luu, J. X., Marsden, B., Jewitt, D., 
Trujillo, C. A., Hergenother, C. W., Chen, J., \& Offutt, W. B. 1997,
Nature, 387, 573

\bibitem[Malhotra 1993]{mal93} Malhotra, R. 1993, Nature, 365, 819

\bibitem[Malhotra 1995]{mal95} Malhotra, R. 1995, AJ, 110, 420

\bibitem[Marzari \etal 1997]{mar97} Marzari, F., Farinella, P., Davis, D. R.,
Scholl, H., Bagatin, A. C. 1997, Icarus, 125, 39

\bibitem[Morbidelli \& Valsecchi 1997]{mor97} Morbidelli, A., \& 
Valsecchi, G. B. 1997, Icarus, 128, 464

\bibitem[Ohtsuki \& Nakagawa 1988]{oht88} Ohtsuki, K., \& 
Nakagawa, Y. 1988, Prog Theor Phys (Suppl), 96, 239

\bibitem[Ohtsuki \etal 1990]{oht90} Ohtsuki, K., Nakagawa, Y., \&
Nakazawa, K. 1990, Icarus, 83, 205

\bibitem[Osterloh \& Beckwith 1994]{ost94} Osterloh, M. \& Beckwith, 
S. V. W. 1994, ApJ, 439, 288

\bibitem[Pollack 1984]{pol84} Pollack, J. B. 1984, ARA\&A, 22, 389

\bibitem[Pollack \etal 1996]{pol96} Pollack, J. B., Hubickyj, O., 
Bodenheimer, P., Lissauer, J. J., Podolak, M., \& Greenzweig, Y. 1996,
Icarus, 124, 62

\bibitem[Roddier \etal 1996]{rod96} Roddier, C., Roddier, F.,
Northcott, M. J., Graves, J. E. \& Jim, K. 1996, ApJ, 463, 326

\bibitem[Roques \etal 1994]{roq94} Roques, F., Scholl, H., Sicardy, B.,
\& Smith, B. A. 1994, Icarus, 108, 37

\bibitem[Russell \etal 1996]{rus96} Russell, S. S., Srinivasan, G., 
Huss, G. R., Wasserburg, G. J., \& Macpherson, G. J. 1996, Science, 273, 757

\bibitem[Ryan \& Melosh 1998]{rya98} Ryan, E. V., \& Melosh, H. J. 1998,
Icarus, 133, 1

\bibitem[Safronov 1969]{saf69} Safronov, V. S. 1969, Evolution of
the Protoplanetary Cloud and Formation of the Earth and Planets,
Nauka, Moscow [Translation 1972, NASA TT F-677]

\bibitem[Sargent \& Beckwith 1993]{sar93}
Sargent, A. I., \& Beckwith, S. V. W. 1993, Phy Tod, 46, 22

\bibitem[Spaute \etal 1991]{spa91} Spaute, D., Weidenschilling, S. J.,
Davis, D. R., \& Marzari, F. 1991, Icarus, 92, 147 

\bibitem[Stapelfeldt \etal 1998]{sta98} Stapelfeldt, K. R., Krist, J. E.,
M\'enard, F., Bouvier, J., Padgett, D. K., \& Burrows, C. J. 1998, ApJ,
502, L65

\bibitem[Stern 1995]{ste95} Stern, S. A. 1995, AJ, 110, 856

\bibitem[Stern 1996]{ste96} Stern, S. A. 1996, AJ, 112, 1203

\bibitem[Stern \& Colwell 1997a]{st97a} Stern, S. A., \& 
Colwell, J. E. 1997a,  AJ, 114, 841

\bibitem[Stern \& Colwell 1997b]{st97b} Stern, S. A., \& 
Colwell, J. E. 1997b,  ApJ, 490, 879

\bibitem[Tremaine 1990]{tre90} Tremaine, S. 1990, In {\it Baryonic Dark 
Matter,} edited by D. Lynden-Bell \& G. Gilmore, Kluwer, Dordrecht, p. 37

\bibitem[Ward \& Hahn 1998]{war98} Ward, W. R., \& Hahn, J. M. 
1998, AJ, 116, 489

\bibitem[Weidenschilling 1977]{wei77} Weidenschilling, S. J. 1977,
Astrophys Sp Sci, 51, 153

\bibitem[Weidenschilling 1997]{we97a} Weidenschilling, S. J. 1997,
Icarus, 127, 290

\bibitem[Weidenschilling \etal 1997]{we97b} Weidenschilling, S. J., 
Spaute, D., Davis, D. R., Marzari, F., \& Ohtsuki, K. 1997, Icarus, 128, 429

\bibitem[Wetherill 1990]{wet90} Wetherill, G. W. 1990,
Icarus, 88, 336

\bibitem[Wetherill \& Stewart 1989]{wet89}
Wetherill, G. W., \& Stewart, G. R. 1989.  Icarus 77, 300 - 357. 

\bibitem[WS93]{ws93} Wetherill, G. W., \& Stewart, G. R. 1993,  
Icarus, 106, 190 (WS93) 

\bibitem[Williams \& Wetherill 1994]{wil94} Williams, D. R., \&
Wetherill, G. W. 1994, Icarus, 107, 117

\bibitem[Williams \etal 1995]{wil95} Williams, I. P., O'Ceallaigh, D. P.,
Fitzsimmons, A., \& Marsden, B. G. 1995, Icarus, 116, 180

\bibitem[Wurm \& Blum 1998]{wur98} Wurm, G., \& Blum, J. 1998, Icarus, 132, 125

\end{thebibliography}
\end{document}